\documentclass[preprint,eqsecnum,flushrt]{aastex}

\usepackage{natbib}
\usepackage{graphicx}

\usepackage{amsmath, amsthm, amssymb }

\newcommand{\realTemp}{\mbox{$1$} MK}
\newcommand{\realnumdens}{\mbox{$10^{8}$} cm\textsuperscript{-3}}
\newcommand{\realB}{\mbox{$10$} G}
\newcommand{\realbeta}{\mbox{$0.007$}}
\newcommand{\realL}{\mbox{$100$} Mm}
\newcommand{\realvao}{\mbox{$2.2$} Mm/s}
\newcommand{\realTao}{\mbox{$290$} MK}
\newcommand{\realReynolds}{\mbox{$2982$}}
\newcommand{\PrandtlSim}{\mbox{$0.01$}}
\newcommand{\realalfvtime}{\mbox{$46$} s}

\newcommand{\timeSim}{\mbox{$2550$} s}

\newcommand{\doublezeta}{\mbox{$90$} degrees}
\newcommand{\numsegments}{\mbox{$4000$}}
\newcommand{\fintime}{\mbox{$3200$} s}

\newcommand{\MachSim}{\mbox{$3.85$}}
\newcommand{\MachSimShock}{\mbox{$5.3$}}
\newcommand{\LSubShock}{\mbox{$22$} Mm}
\newcommand{\SStime}{\mbox{$532$} s}
\newcommand{\LHeatFront}{\mbox{$93$} Mm}
\newcommand{\HHtime}{\mbox{$188$} s}
\newcommand{\LHeatFrontAlfven}{\mbox{$28.0$} Mm}
\newcommand{\elecmfpunitless}{\mbox{$0.0021$}}
\newcommand{\protmfpunitless}{\mbox{$0.9 \times 10^{-3}$}}

\newcommand{\PrandtlSuperCrit}{ \mbox{$0.01,0.03,0.1,0.3,1.0$}, and \mbox{$3.0$}}

\begin{document}

\title{Shocks and Thermal Conduction Fronts in Retracting Reconnected Flux Tubes}

\author{S. E. Guidoni, D. W. Longcope}

\affil{Department of Physics, Montana State University, Bozeman, MT 59717-3840} 
\email{guidoni@physics.montana.edu}

\begin{abstract}

We present a model for plasma heating produced by time-dependent, spatially localized reconnection within a flare current sheet separating skewed magnetic fields. The reconnection creates flux tubes of new connectivity which subsequently retract at Alfv\'enic speeds from the reconnection site.  Heating occurs in gas-dynamic shocks which develop inside these tubes. Here we present generalized thin flux tube equations for the dynamics of reconnected flux tubes, including
pressure-driven parallel dynamics as well as temperature
dependent, anisotropic viscosity and thermal conductivity. The
evolution of tubes embedded in a uniform, skewed magnetic field,
following reconnection in a patch, is studied through numerical
solutions of these equations, for solar coronal conditions. Even though viscosity and thermal
conductivity are negligible in the quiet solar corona, the strong
gas-dynamic shocks generated by compressing plasma inside
reconnected flux tubes generate large velocity and
temperature gradients along the tube, rendering the diffusive
processes dominant. They determine the thickness of the
shock that evolves up to a steady-state value, although this
condition may not be reached in the short times involved in a
flare. For realistic solar coronal parameters, this steady-state
shock thickness might be as long as the entire flux tube. 
For strong shocks at low Prandtl numbers, typical of the solar
corona, the gas-dynamic shock consists
of an isothermal sub-shock where all the compression and cooling
occur, preceded by a thermal front where the temperature
increases and most of the heating occurs. We estimate the length
of each of these sub-regions and the speed of their propagation. 

\end{abstract}


%
\section{Introduction}

Magnetic reconnection has long been accepted as a significant mechanism for energy release in flares \citep{Kopp_1976,Priest_1981}, coronal mass ejections \citep{Forbes_2000,Klimchuk_2001,Low_2001,Lin_2003}, and as an interaction process between the solar wind and the magnetosphere \citep{Pudovkin_1985}. 
It operates in current sheets across which there is a discontinuity in the direction of the pre-reconnection field lines. Some kind of diffusion mechanism initiates the connection of field lines from one side of the current sheet to field lines on the other side. The exact nature of this process is not yet well understood, but it has recently become clear that it must produce an electric field restricted to a small portion of the current sheet \citep{Biskamp_2001,Birn_2001}. Magnetic reconnection was first modeled in two-dimensions and steady state with uniform resistivity \citep{Sweet_1958,Parker_1957}, but the energy release in this case was too slow to explain the rapid development observed in flares and CMEs \citep{Biskamp_1986}. This problem was solved by assuming a 2D steady electric field {\em localized} to a small diffusion region with four slow-mode shocks (SMSs) attached to it. At these shocks, the magnetic field is deflected and partially annihilated, and the plasma is heated and accelerated \citep{Petschek_1964}. The compressible plasma case was studied by \citet{Soward_1982}. 

Localized, {\em non-steady} models, began with the work of \citet{Semenov_1983}, followed by \citet{Biernat_1987} and \citet{Nitta_2002} all of which were two-dimensional.  For these spatially localized (in only one dimension) and short lived episodes, four slow mode shocks (SMSs) extend from the reconnection site as well, but they close back together in a finite region that increases in size as time goes by (as opposed to the infinite length SMSs from steady state). These teardrop-shaped SMSs continue growing and retracting as they sweep up mass, even if reconnection ceases \citep{Semenov_1998}. Therefore, this conversion of magnetic energy into kinetic and thermal energy continues and the dissipated energy can exceed the ohmic dissipation from the reconnection electric field, indicating that a small scale, short lived event can affect the global structure of magnetic fields. 

There is some observational evidence supporting the basic features of fast magnetic reconnection models.  The heating and chromospheric evaporation in flares provides indirect evidence of the formation and propagation of slow mode shocks \citep{Lin_2008}. Many observational features, as plasmoids moving along current sheets can be attributed to unsteady, Petschek-type reconnection, although they also can be the result of turbulent reconnection \citep{Lin_2008}.  Current sheets themselves are notoriously difficult to observe directly, yet they are a necessary component of all fast magnetic reconnection models.  Those observations providing evidence \citep{Schettino_2010} also suggest high temperatures at the current sheet, consistent with shock heating of material.

There is also evidence for localized and transient nature of magnetic reconnection.
 The first such evidence was found in so called {\em flux transfer events} identified 
 in {\em in situ} magnetospheric observations \citep{Russell_1978}. 
X-ray and EUV observation have recently provided evidence of solar reconnection episodes localized in space and time. Dark voids \citep{McKenzie_1999} descending from reconnection regions are directly related to the path of bright loops that appear lower down in the corona \citep{Sheeley_2004}.
These downward-moving features are believed to represent flux tubes, created by patchy reconnection, retracting downward to later form the post-eruption arcade \citep{McKenzie_2009}. 

All of the models above considered reconnection between magnetic field lines assumed to be perfectly antiparallel. A departure from this special case, called reconnection between 
{\em skewed} fields, introduces a magnetic field component in the ignorable direction, called a guide-field; such models are sometimes called 
2.5-dimensional.
In this case it is necessary to supplement the SMSs with rotational discontinuities 
\citep[RDs,][]{Petschek_1967,Soward_1982_II,Heyn_1988}.  The effect of the RDs is to produce bulk flow in the ignorable direction in order that the velocity change, magnetic fields and shock normal may all lie in a single plane; this is a requirement of the SMS often called ``co-planarity''.  This new flow component is confined to the region between the RD and SMS.
It is the RDs rather than the SMSs which accelerate the reconnection flux to the Alfv\'en speed in an outflow jet.

Non-steady reconnection between skewed magnetic fields was analyzed by \citet{Semenov_1992} and \citet{Biernat_1998} for dayside magnetospheric events. 
\citet{Linton_2006} studied the relaxation of a solar-like low plasma-$\beta$ reconnected flux tube via three-dimensional magnetohydrodynamic simulations. Even though the initial configuration is 2.5-dimensional, reconnection is initiated within a small patch, producing a distinct bundle of reconnected field lines, a flux tube; the ensuing reconnection dynamics is purely three-dimensional. They found that the perpendicular dynamics of the reconnected tube was well described by equations for thin flux 
tubes \citep[TFT,][]{Spruit_1981}.  The equations differed from those used traditionally in convection zone models, by their application to the low-$\beta$ corona.  Linton and Longcope's study shows RDs propagating along the legs of the tube that differ from the usual Petschek SMSs in the sense that they rotate the magnetic field without changing its magnitude. 
In this respect, they were like the RDs in steady-state 2.5-dimensional models 
\citep{Soward_1982_II,Skender_2003}.  Rather than a coherent outflow jet, 
the Alfv\'en-speed retraction of reconnected flux occurs as a single, transient  event.

Common to all models of fast magnetic reconnection, except those restricted to perfectly anti-parallel fields, is a second shock where plasma heating occurs, located inside the RDs.   In the steady models of \citet{Soward_1982_II} and \citet{Skender_2003}, this inner shock is the SMS.  At low plasma $\beta$ and skew angles significantly far from anti-parallel, the SMS differs significantly from a switch-off shock: it changes the magnetic field strength very little and derives most of its thermal energy from bulk flow parallel to the field lines.  Recently, \citet{Longcope_2009} presented a TFT model which included flow parallel to the field lines \citep[had not  investigated the parallel plasma dynamics]{Linton_2006}.  They found a shock of apparently different character, a gas dynamic shock (GDS), changing only the parallel flow and leaving the magnetic field strength completely unchanged.  Upon further consideration, however, this shock appears related to the nearly-parallel SMS in 2.5 steady models.  In fact the post-GDS temperatures in the TFT model are almost identical to the post-SMS temperatures in steady-state MHD models.  In both kinds of models, plasma heating occurs through a two-stage process notably different from commonly-used switch-off models, applicable to purely anti-parallel reconnection.  First the RD converts magnetic energy to bulk kinetic energy in the form of Alfv\'en-speed flow partly {\em parallel} to the field line.  The parallel component of this flow, present in both, is then thermalized in a second shock of large {\em hydrodynamic} Mach number.

Also common to all the reconnection models is the presence of Alfv\'enic outflows originating at the reconnection site.  In steady state models these form coherent jets, for which there is some observational support, although it is typically indirect \citep{Tsuneta_1997_I}.  In transient reconnection models any outflow is more accurately described as a distinct retraction event rather than a steady jet.  Observational evidence for these events is provided by the ongoing investigations of downflowing features above flare arcades 
\citep{McKenzie_1999,Sheeley_2004,McKenzie_2009}.  While few if any of these events have measured  speeds matching the local Alfv\'en speed, they are generally interpreted as evidence of reconnection. Either these outflows/downflows were Alfv\'enic at some earlier time or all reconnection theories must be revised.  In the model presented here we continue to assume reconnection outflow at the Alfv\'en speed.

While all of the foregoing reconnection models predict shocks, few investigations of them have explicitly included viscosity or thermal conductivity.  This omission is noteworthy since strictly speaking the irreversible nature of any shock demands a non-ideal effect; in the collisional plasma of the solar corona this is usually assumed to be viscosity.
Nevertheless reconnection models generally neglect viscosity altogether to say nothing of accounting for its strong temperature dependence and directional anisotropy.  \citet{Craig_2005} and \citet{Litvinenko_2005} studied the effects of temperature independent viscosity in various reconnection disturbances. Only recently, \citet{Craig_2008} and \citet{Craig_2009} included anisotropic and temperature dependent viscous dissipation in a three-dimensional steady-state model at X-points. 

For strongly magnetized plasma, viscous diffusion and thermal conduction are highly anisotropic.  They are directed along the magnetic field since perpendicular coefficient are many orders of magnitude smaller than their parallel counterparts.  
Transport coefficients in a fully-ionized plasma each have a strong dependence on  temperature ($\sim T^{\frac{5}{2}}$) making transport far more significant at high post-shock temperatures than it is assumed to be in general.  It has therefore been recognized as a potentially significant factor in flares  \citep{Cargill_1995,Forbes_1986}.  In particular the heat generated at SMSs is predicted to be conducted along reconnected field lines ahead of other reconnection effects, possibly driving chromospheric evaporation \citep{Forbes_1989}. 
Yokoyama \& Shibata (1997, 2001) included anisotropic thermal conduction in 2D simulations of reconnection, and \citet{Seaton_2009} presented a detailed theoretical analysis of these simulations. In most of these two-dimensional, anti-parallel models, the heat conduction front extends in front of SMSs as expected.

Here we seek to understand the role played by anisotropic transport in transient (non-steady) reconnection between skewed magnetic fields.  Toward this end we generalize the TFT equations of \citet{Linton_2006} and \citet{Longcope_2009} to
include pressure-driven parallel dynamics as well as temperature dependent, anisotropic viscosity and thermal conductivity.  The latter are essential for self-consistently producing the shocks predicted by \citet{Longcope_2009} analytically using conservation laws.


After presenting the general equations, we consider the flare-like scenario depicted in Fig. \ref{fig:flare_CS} where reconnection occurs within a current sheet above an arcade of post-flare loops.  Short lived, patchy (fast) reconnection (depicted in the Figure as a small sphere) is assumed to reconnect the skewed field lines from opposite sides of the sheet. The small patch creates a small bundle of field lines that subsequently retract at Alfv\'enic speeds. We approximate the magnitude of the magnetic field as uniform and assume strongly magnetized plasma (low plasma-$\beta$). With these assumptions, modified thin flux tube equations are presented and the evolution of a thin flux tube that has been reconnected is analyzed in detail.

We solve our new equations numerically for the same reconnection scenario studied by \citet{Longcope_2009}, using realistic solar coronal parameters. The diffusive processes
determine the thickness and internal structure of the shocks.
We analyze this internal structure using methods similar to those of \citet{Thomas_1944}, \citet{Grad_1952}, \citet{Gilbarg_1953}, and 
\citet{Kennel_1988}.  Our own analysis is novel in its focus on solar flare conditions, including the very low Prandtl number (ratio of viscosity to thermal conductivity) and strong temperature dependence.   
In particular we find that for strong shocks at low Prandtl number, the gas dynamic shock comprises a narrow iso-thermal sub-shock behind a very large conduction front.  Due to the strong temperature dependence of the coefficients the conduction front has significant extent; far greater than the term ``internal structure'' might suggest.  It is related to the conduction front anticipated in two-dimensional anti-parallel models, but here we offer a detailed analysis of its structure.

In addition to thermal conductivity we find it it is necessary to include viscosity in order to account for the isothermal sub-shock region.
Most of the plasma compression occurs isothermally within the sub-shock so the plasma is actually cooled by it. \citet{Fowles_1975} showed that for super-critical shocks, it is impossible to assume inviscid behavior, no matter how large thermal conduction is. \citet{Coroniti_1970} studied dissipation discontinuities in shock waves for temperature independent transport coefficients.  The seemingly harmless assumption of temperature independent coefficients leads to shock thicknesses on the order of the particle's mean-free-path, raising concerns about the adequacy of a fluid treatment: when the thickness of the shock is of the order of particle mean free-path, the fluid model breaks down \citep{Campbell_1984}. When transport coefficients are given an accurate temperature dependence, however, we find the thickness of the shocks exceeds the particle mean free paths by at least an order of magnitude.  Our simulations also show that the ratio between the heat flux and the nominal free-streaming heat flux is smaller than $0.004$. These two results allow us to treat our plasma as a collisional fluid. 

The layout of the present paper is the following. The next section is a brief review of the general transport effects in magnetohydrodynamics. The following section describes the thin flux tube assumptions that lead to the final thin flux tube equations presented in the third section. Section \ref{sec:sims} presents simulations of the retraction of a thin flux tube reconnected in a localized region inside a current sheet with uniform, skewed magnetic fields for a case with low plasma-$\beta$, and parameters relevant to the solar corona. In the fifth section, an analytical analysis of the inner structure of the shocks is presented. Finally, in the last section, the main results of the paper and possible observational implications of the inner structure of the shocks are discussed.  

%
%
%
\section{MHD equations}
   \label{sec:MHD_equations}

When resistivity, radiation and gravity are neglected, the magneto-hydrodynamic (MHD) equations for a charge-neutral plasma (we will assume sufficient collisionality to justify their use) are \citep[chap. 22]{Shu_VolII}:
%
%
\begin{eqnarray}
   \label{eqn:MHD_mass}
      \frac{1}{\rho} \frac{D\rho}{Dt} & = & - \nabla \cdot \mathbf{v}, \\
   \label{eqn:MHD_momentum}
       \rho \frac{D\mathbf{v}}{Dt} & = & 
      -\nabla \left(P + \frac{B^{2}}{8\pi}\right)+ \nabla \cdot 
      \left(\frac{\mathbf{B}\mathbf{B}}{4 \pi}\right)-\nabla \cdot \Pi,  \\
   \label{eqn:MHD_energy}
      \rho T \frac{Ds}{Dt} & = & \rho \frac{D\varepsilon_{I}}{Dt} - \frac{P}{\rho }\frac{D \rho}{Dt} = \Psi -\nabla \cdot \mathbf{q}, \\
   \label{eqn:MHD_induction}
      \frac{\partial \mathbf{B}}{\partial t}
      & = & - \nabla \times \left(\mathbf{B} \times \mathbf{v}\right).
\end{eqnarray}
%
%
\noindent Here, $\rho$, $P$, $\mathbf{B}$, $\mathbf{v}$, $T$, $s$, $\varepsilon_{I}$, $\Pi$, $\Psi$, and $\mathbf{q}$ are the density, plasma pressure, magnetic field, velocity, temperature, specific entropy, specific internal energy, viscous stress tensor, viscous heat, and thermal conduction flux of the fluid, respectively.
The advective derivative is defined as ${D}/{Dt} =
{\partial}/{\partial t}+\mathbf{v}\cdot\nabla$. 
In Eq. \eqref{eqn:MHD_momentum}, dyadic notation \citep[chap. 5]{Goldstein_Cl_Mech} is used for the second right-hand-term.


When the plasma is strongly magnetized by a field taken in the $\mathbf{\widehat{z}}$ direction, the stress tensor takes a simple gyrotropic form \citep{Braginskii_1965} 
%
%
\begin{eqnarray}
   \label{eqn:MHDstresst}
      \Pi &=& -\eta W_{zz} \left(-\frac{\mathbf{\widehat{x}}\mathbf{\widehat{x}}}{2}-
      \frac{\mathbf{\widehat{y}}\mathbf{\widehat{y}}}{2}
       +\mathbf{\widehat{z}}\mathbf{\widehat{z}}\right), 
\end{eqnarray}
where the $\mathbf{\widehat{z}}\mathbf{\widehat{z}}$ component of the rate-of-stress tensor is
\begin{eqnarray}
   \label{eqn:MHD_rate_of_strebght}
       W_{zz} & = & 2 \frac{\partial v_{
       z}}{\partial z} - 
       \frac{2}{3}\nabla \cdot \mathbf{v}.
\end{eqnarray}
%
%
\noindent Here, $\eta$ is the largest of the dynamic viscosity coefficients, determined essentially by the
ions (the electron-to-ion viscosity ratio is ${\eta_{e}}/{\eta_{i}} \simeq$ \PrandtlSim). 
%
The strongly magnetized limit of the conductive heat flux is
%
%
\begin{eqnarray}
   \label{eqn:MHD_qflux} \mathbf{q} & = & 
      - \kappa \nabla_{\parallel}(k_{B} T).
\end{eqnarray}
%
%
\noindent where $\nabla_{\parallel} = \mathbf{\widehat{z}}\mathbf{\widehat{z}} \cdot \nabla$  is the derivative parallel to the magnetic field, and $\kappa$ is the parallel thermal conduction coefficient, determined mostly by the electrons (the electron-to-ion heat conductivity ratio, 
${\kappa_{e}}/{\kappa_{i}} \simeq 24$). 
%

We close the system using the ideal gas law
%
%
\begin{eqnarray}
   \label{eqn:press_gas}
      P & = & \frac{\rho k_{B} T}{\overline{m}} = 2 n k_{B} T, \\
    \label{eqn:energy_gas}
       \varepsilon_{I} & = & \frac{P}{(\gamma-1) \rho}, \\
   \label{eqn:entropy_gas}
      s & = & c_{V} \ln \left[ \frac{P}{\rho^{\gamma}} \right] + s_{0},
\end{eqnarray}
%
where $\gamma$ is the adiabatic gas constant, $\overline{m}$ the average particle mass (for a fully ionized Hydrogen plasma, it is approximately equal to half the mass of the proton), $n$ is the electron number density, $c_{V}= {k_{B}}/{(\gamma-1) \overline{m}} $ is the specific heat at constant volume, and $s_{0}$ is a constant.

%
%
%
%
%
%

\section{Thin Flux Tube Equations}

\subsection{The Thin Flux Tube Assumptions}

\begin{figure}
   \centering
   \includegraphics{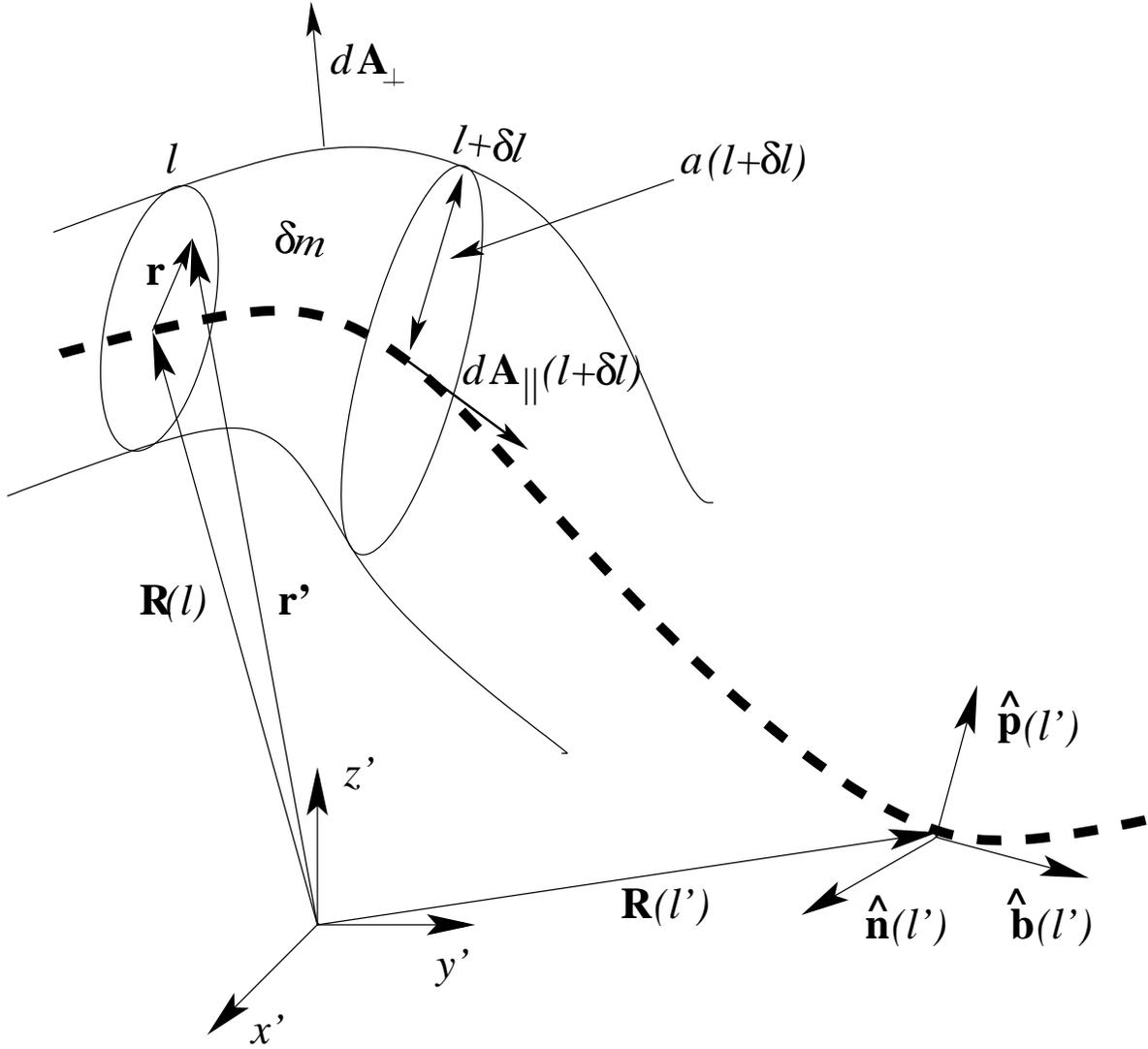}
   \caption
   {
      Frenet coordinate system. The axis of a thin magnetic flux tube is represented by a central magnetic field line (dashed line) parameterized by vector $\mathbf{R}(l)$. The unit vector
      $\mathbf{\widehat{b}}$ is parallel to the local magnetic field,
      unit vector $\mathbf{\widehat{p}}$ points in the direction of the
      curvature vector $\mathbf{k}$, and unit vector
      $\mathbf{\widehat{n}}$ completes the system. The two ellipses are the cups of a small tube segment of mass $\delta \hbox{m}$. 
   }
   \label{fig:frenetsystem} 
\end{figure}

We consider an untwisted, \textit{thin} tube of magnetic flux $\phi$, like the one depicted in Fig. \ref{fig:frenetsystem}, embedded in an strongly magnetized external fluid. The external magnetic field, thermal pressure and density will be denoted $\mathbf{B}_{e}$, $P_{e}$, $\rho_{e}$, respectively. These quantities may depend on position, but not time, since we will assume the background configuration to be in equilibrium. The tube is considered so thin that the background quantities remain unperturbed as the tube moves through the external plasma. 
It is also tine compared to the local radius of curvature.

We will assume that the tube can be represented by an internal field 
line, $\mathbf{R}(l)$, the axis (see Fig. \ref{fig:frenetsystem}).  The field is untwisted in the sense that all field lines are parallel to the axis.
The time it takes for a fast magneto-sonic wave to establish pressure balance with its surroundings scales as $\tau_{fm}={a(l)}/{\sqrt{c_{s}^{2}(l) + v_{a}^{2}(l)} }$, where $v_{a}$ is the Alfv\'{e}n speed, and $c_{s}$ is the sound speed. Since the fast-magnetosonic time is negligible with respect to the Alfv\'en time over a characteristic scale $L_{ch}$, 
$\tau_{A} = {L_{ch}}/{v_{a}(l)}$, we assume instantaneous, local pressure balance
%
%
\begin{eqnarray}
   \label{eqn:TP_constrain}
       P + \frac{B^{2}}{8 \pi} & = & P_{e} + \frac{B^{2}_{e}}{8 \pi},
\end{eqnarray}
%
\noindent where, $P$ and $B$ are pressure and magnetic field inside the tube. 

It is convenient to define a Frenet coordinate about the axis $\mathbf{R}$,  parameterized by its arc-length $l$, as shown in Fig. \ref{fig:frenetsystem}.  The Frenet unit vectors defining the coordinate system are
%
%
\begin{eqnarray}
   \label{eqn:bvect}
      \mathbf{\widehat{b}}(l)&=&\frac{\mathbf{B}[\mathbf{R}(l)]}{B[\mathbf{R}(l)]}
      = \frac{\partial }{\partial l}\mathbf{R}(l), \\
   \label{eqn:kvect} 
      \mathbf{\widehat{p}}(l)&=&\mathbf{k}(l)r_{c}(l)=\frac{\partial \mathbf{\widehat{b}}(l)}{\partial
      l}r_{c}(l), \\
   \label{eqn:nvect}
       \mathbf{\widehat{n}}(l) &=&
       \mathbf{\widehat{p}}(l) \times \mathbf{\widehat{b}}(l).
\end{eqnarray}
%
%
\noindent The unit vector $\mathbf{\widehat{b}} $ is parallel to the
axis, $\mathbf{\widehat{p}}$ is the normal vector and $\mathbf{\widehat{n}}$ is the 
bi-normal completing the system. 
Local derivative operators associated with the unit vectors are
%
%
\begin{eqnarray}
   \label{eqn:dirder}
      \partial_{\parallel} & = & \widehat{\mathbf{b}} \cdot \nabla = \partial_{l}, \nonumber \\
      \nabla_{\bot}  & = & - \widehat{\mathbf{b}} \times (\widehat{\mathbf{b}} \times\nabla).  \nonumber
\end{eqnarray}
%


%
%
We assume the flux tube is sufficiently isolated from its surrounding that no viscous stress crosses its surface.  The stress tensor is therefore
\begin{eqnarray}
   \label{eqn:tft_str_tens}
      \Pi &=& -\eta \left( \widehat{\mathbf{b}} \cdot \frac{\partial \mathbf{v}}{\partial l}  \right)       \widehat{\mathbf{b}}\widehat{\mathbf{b}}. 
\end{eqnarray}
%
For parameters typical at the high corona, T = \realTemp, n = \realnumdens, B = \realB, L = \realL, the Reynold's number is $R_{\eta} \simeq 3 \times 10^{3}$, indicating a relative small contribution from viscosity. 
For such high Reynolds numbers, the viscous stress tensor may become significant only at steep velocity gradients, like shocks. Shocks in thin flux tubes were predicted by 
\citep{Longcope_2009}. 

The small Prandtl number of an ionized plasma, 
$P_{r} = \eta / \kappa \overline{m} \sim $ \PrandtlSim, indicates that the effect of thermal conduction is larger than viscosity. Temperature gradients will then be even smaller at shocks than velocity gradients.

%
%
%

\subsection{Thin Flux Tube Equations}

The thin flux tube equations are derived by applying the MHD equations to a small piece of the tube with mass $\delta \hbox{m}$, like the one shown in Fig. \ref{fig:frenetsystem}. The position of every point in the tube can be expressed as $\mathbf{r'} = \mathbf{R} + \mathbf{r}$. Vector $\mathbf{r}$ belongs to the circular area perpendicular to the tube axis, and can be expressed as $\mathbf{r}=r \sin \theta \hbox{ }\mathbf{\widehat{n}} + r \cos \theta \hbox{ }\widehat{\mathbf{p}} =  r\mathbf{\widehat{r}}$, where $r$ and $\theta$ are the polar coordinates associated to the $\mathbf{\widehat{n}}$-$\mathbf{\widehat{p}}$ plane. 

The local radius of the tube, $a(l)$, is small compared with scales of variation along the tube, permiting Taylor-expand any state variable $f$ as follows
%
%
\begin{eqnarray}
\label{eqn:expansion}
   f(\mathbf{r}) & = & \mathop{\sum_{k=0}^{\infty}}
   \frac{\left[(\mathbf{r}\cdot\nabla)^{k}f\right](l)}{k!} = \mathop{\sum_{k=0}^{\infty}} \mathop{\sum_{i=0}^{k}} \binom{k}{i}
      \frac{r^{k}}{k!}(\cos \theta)^{k-i}
      (\sin \theta)^{i} \frac{\partial^{k}f(l)}{\partial
      n^{k-i}\partial p^{i}}.
\end{eqnarray}
%
%
The second equality was obtained expanding the binomial term in the local polar coordinates.
Thinness also allows us to assume that the piece's axis lays on a plane, and its radius of curvature is constant along the small piece. For this case, the differentials of area and volume for the piece can be expressed as
%
%
\begin{eqnarray}
    d\mathbf{A_{\parallel}} & = &
    \widehat{\mathbf{b}}(l) \hbox{ }r\hbox{ } dr\hbox{ } d\theta,    \\
    d\mathbf{A_{\perp}}  & = &  \left.\left[ \frac{\partial
    \mathbf{r'}}{\partial \theta} \times \frac{\partial
    \mathbf{r'}}{\partial l}\right]\right|_{r=a(l)} dl \hbox{ } d\theta, \nonumber \\
    & = &  \left\{\mathbf{\widehat{r}}(l) \left[a(l)-
   \frac{a^{2}(l)\sin\theta}{r_{c}(l)} \right]  -
   \widehat{\mathbf{b}}(l)  a(l)\frac{\partial a(l)
   }{\partial l} \right\} dl d\theta  ,     \\
   d\hbox{\textit{v}}  & = & \left[ \frac{\partial \mathbf{r'}}{\partial
   \theta} \times \frac{\partial \mathbf{r'}}{\partial
   l}\right]\cdot \frac{\partial
   \mathbf{r'}}{\partial r}\hbox{ }  \hbox{ } dl \hbox{ }d\theta \hbox{ } dr \nonumber \\
   \label{eqn:diff_vol}
      & = & \left(r-\frac{r^{2} \hbox{ } \sin\theta}{r_{c}(l)}\right)
      \hbox{ }  \hbox{ } dl \hbox{ }d\theta \hbox{ } dr,   
\end{eqnarray}
%
%
 
An example of the expansion \eqref{eqn:expansion} is the mass of the tube piece. Computing the volume integral using expression \eqref{eqn:diff_vol} for the volume differential, and expanding the density around the center of the tube, gives
%
%
\begin{eqnarray}
   \delta \hbox{m} & = & \int \rho \hbox{ } d\hbox{v} \nonumber \\
   & = & \mathop{\sum_{k=0}^{\infty}}
   \mathop{\sum_{i=0}^{k}} \binom{k}{i} \frac{1}{k!} \int_{l }^{l+
   \delta l} \frac{\partial^{k}\rho(l')}{\partial n^{k-i}\partial
   p^{i}}\hbox{ }dl' \int_{0}^{a(l)} \int_{0}^{2\pi}
   r^{k+1} \times \nonumber \\
   &  &  \times \hbox{ } (\cos \theta)^{k-i} (\sin \theta)^{i}  \hbox{ } \left(1-\frac{r \hbox{ }
   \sin\theta}{r_{c}(l)}\right)  d\theta dr.  \nonumber \\
   & = &  \int_{l }^{l+ \delta l} \rho(l') \pi a^{2}(l') \hbox{ }dl'
   + \int_{l }^{l+ \delta l} \mathcal{O}(a^{4}(l')) dl', \nonumber
\end{eqnarray}
%
%
where $\mathcal{O}(a^{4}(l'))$ represents an error of fourth order
in the local radius. If the arc-length interval, $\delta l$, is small then the above equation can be expressed as
%
%
\begin{eqnarray}
   \label{eqn:delta_mass}
      \delta \hbox{m}(l) & \simeq & \rho(l) \pi a^{2}(l) \delta l,
\end{eqnarray}
%
%
Expanding the dot product $\mathbf{B} \cdot \widehat{\mathbf{b}}$, gives a similar expression for the magnetic flux
%
%
\begin{eqnarray}
   \label{eqn:flux_tft}
      \phi & =& \int_{l} \mathbf{B}\cdot d\mathbf{A_{\parallel}} \nonumber \\
      & =& B(l) \pi a^{2}(l) +  \mathcal{O}(a^{4}(l)) \simeq B(l) \pi a^{2}(l) . 
\end{eqnarray}
%
%
%

For a given tube piece, the ratio between the mass and the magnetic flux, $\delta \mu  = \delta \hbox{m}/\phi = (\rho/B) \delta l$, is constant. This definition, along with Eq. \eqref{eqn:delta_mass} and \eqref{eqn:flux_tft}, allow us to define the integrated mass per flux, 
$\mu = \int{(\rho/B) dl } $, to parameterize the tube instead of $l$.
The differential with respect to $\mu$ is
%
%
\begin{eqnarray}
   \label{eqn:deriv_mu}
   \frac{\partial }{\partial \mu}  & =& \frac{B}{\rho} \frac{\partial}{\partial l}.
\end{eqnarray}
%
%

In order to obtain the thin flux tube mass equation we integrate over the piece's volume every term of the corresponding MHD equation, and keep only the first non-vanishing terms. 
To first order, the velocity at any point in the tube is the sum of the velocity of the center of the tube plus a radial term $v_{r}$. This radial velocity at the surface of the tube is equal to the change in local radius $(D/Dt) a(l)$. Finally, applying the fundamental theorem of calculus to recover integrals along the arc-length, and expression \eqref{eqn:flux_tft}, the final result is
%
\begin{eqnarray}
   \label{eqn:TFT_mass} 
      \frac{D}{Dt}\left(\frac{B}{\rho}\right) & = & \widehat{\mathbf{b}} \cdot \frac{\partial \mathbf{v}}{\partial \mu}
\end{eqnarray}
%
%

In the integration of the momentum equation, \eqref{eqn:MHD_momentum},
the surface integrals for the perpendicular area vanish since $\mathbf{B} \cdot \mathbf{dA_{\perp}} = 0$ and $\Pi \cdot \mathbf{dA_{\perp}}=0$ (in this last equality we have used definition \eqref{eqn:tft_str_tens}). For the first term of the right hand side, constraint \eqref{eqn:TP_constrain} is applied and expanded around the center of the tube yielding
%
%
%
\begin{eqnarray}
   \label{eqn:TFT_mom} 
      \frac{D\mathbf{v}}{Dt} & = & \frac{1}{\rho} \left[
      -\widehat{\mathbf{b}} \frac{\partial P}{\partial l}-\nabla_{\perp}
      \left(P_{e} + \frac{B^{2}_{e}}{8 \pi}\right ) +2
      \mathbf{k}\left(P_{e}-P+ \frac{B^{2}_{e}}{8 \pi}\right) \right] + \nonumber \\
      &  & + \frac{B}{\rho} \frac{\partial}{\partial l} \left[ \frac{\widehat{\mathbf{b}\eta}}{B}             \left(\widehat{\mathbf{b}}\cdot \frac{\partial \mathbf{v}}{\partial l}\right)\right],
\end{eqnarray}
%
%
where every term is evaluated at the center of the tube. In the above equation, the viscous momentum term is a perfect derivative in the integrated mass. There is no viscous momentum interchange between the tube and its surroundings.

To determine the form of the viscous heating, we compute the rate of change in kinetic energy as the dot product between the momentum Equation \eqref{eqn:TFT_mom} and the velocity. 
Balancing this loss with a heating rate gives 
%
%
\begin{eqnarray}
   \label{eqn:TFT_visc_heat}
      \Psi & = & \eta
      \left( \widehat{\mathbf{b}} \cdot \frac{\partial \mathbf{v}}{\partial l}\right)^{2}
\end{eqnarray}
%
%
With the above definition, and similar treatment than for the mass and momentum equation, the thin flux tube energy equation becomes
%
%
\begin{eqnarray}
    \label{eqn:TFT_energ}
      \frac{D\varepsilon_{I}}{Dt} & = &  \frac{P}{\rho^{2} }\frac{D \rho}{Dt} + \frac{\eta} {\rho}  \left(\widehat{\mathbf{b}}          \cdot \frac{\partial \mathbf{v}}{\partial l}\right)^{2} +\frac{B}{\rho} \frac{\partial}{\partial l} \left(       \frac{\kappa}{B}  \frac{\partial}{\partial l}(k_{B} T)\right),
\end{eqnarray}
%
%
%
where we see that the thermal conduction term is also a perfect derivative in the integrated mass mass per flux (there is no thermal conduction across the tube).

For our simulations, it is more convenient to re-write the energy equation by defining a new variable $P_{c}$ in the following way
%
%
%
\begin{eqnarray}
    \label{eqn:TFT_P0}
      P(\mu,t) & = &  P_{c}(\mu,t) \left(\frac{\rho(\mu,t)}{\rho(\mu,0) } \right)^{\gamma} = P_{c}(\mu,t) \left(\frac{\rho(\mu,t)}{\rho_e(\mu) } \right)^{\gamma}. 
\end{eqnarray}
%
%
%
%
In an adiabatic case, $P_{c}$ is constant in time. The change in plasma entropy with respect to the initial state (assumed to be in equilibrium) is directly related to $P_{c}$, where $\Delta s = c_{V} \ln \left[{P_{c}(\mu,t)}/{P_{e}(\mu)}\right]$. With definition \eqref{eqn:TFT_P0}, the energy equation is transformed to an equation for $P_{c}$ 
%
%
\begin{eqnarray}
   \label{eqn:TFT_P0_time}
      \frac{D P_{c}}{Dt} & = & (\gamma -1)  \left( \frac{\rho_{e}}{\rho } \right)^{\gamma} 
      \left[ \eta \left(\widehat{\mathbf{b}} \cdot \frac{\partial \mathbf{v}}{\partial l} 
      \right)^{2} + B \frac{\partial}{\partial l} \left( \frac{\kappa}{B}   \frac{\partial}{\partial l}(k_{B} T) 
      \right) \right].
\end{eqnarray}
%
%
%
When density is known, the above equation allows us to update pressure, and use it in the momentum equation. 

Heating and cooling occur where there is a change in entropy. From Eq. \eqref{eqn:MHD_energy}, \eqref{eqn:TFT_energ}, and \eqref{eqn:TFT_P0_time}, the volumetric heating rate can be written as
%
%
\begin{eqnarray}
   \label{eqn:heating}
      \dot{Q} & = & \rho T \frac{Ds}{Dt} =  \frac{1}{\gamma-1} \left( \frac{\rho}{\rho_{e}} \right)^{\gamma}  \frac{D P_{c}}{Dt} \nonumber \\
      & = & \eta \left(\widehat{\mathbf{b}} \cdot \frac{\partial \mathbf{v}}{\partial l} 
      \right)^{2} + B \frac{\partial}{\partial l} \left( \frac{\kappa}{B}   \frac{\partial}{\partial l}(k_{B} T) 
      \right).
\end{eqnarray}
%
%
In this equation, the viscosity term is always positive, and therefore always heats the plasma. On the other hand, the thermal conduction term can have either sign, depending on the temperature's arc-length second derivative. 

The transport coefficients in an ionized plasma depend om temperature as \citep{Spitzer_1962}
%
%
\begin{eqnarray}
   \label{eqn:transp_coeff_eta}
      \eta &=& \eta_{c}T^{\frac{5}{2}}, \\
   \label{eqn:transp_coeff_kappa}
      \kappa &=& \kappa_{c}T^{\frac{5}{2}},
\end{eqnarray}
where $\eta_{c}$ and $\kappa_{c}$ are constants. 

For thin flux tubes, there is no need to solve the induction
equation, since the magnitude of the magnetic field inside the
tube is determined by constraint \eqref{eqn:TP_constrain}, and its direction is given by
the position of the tube piece governed by the momentum equation (resistivity has been neglected, and therefore the magnetic field follows the fluid).

It is possible to describe the TFT equations in dimensionless variables, by defining a characteristic magnetic field  magnitude, $B_{0}$, density, $\rho_{0}$, and length, $L_{0}$. For example, these characteristic variables may represent the magnetic field magnitude and density at the reconnection point, $\mathbf{x}_{R}$, and the length of a typical reconnected tube, respectively. Then, $B_{0} = B_{e}(\mathbf{x}_{R})$, and $\rho_{0}=\rho_{e}(\mathbf{x}_{R})$. With these values we can construct an associated Alfv\'{e}n speed $v_{A0} = B_{0}/ \sqrt{4 \pi \rho_{0}}$, and an associated temperature $T_{0}^{A} = v_{A0}^{2}\overline{m}/ k_{B}$, to non-dimensionalized velocities and temperatures. 

The new dimensionless variables are
%
%
\begin{eqnarray}
    \label{eqn:unitless_var}
       &   \mu^{*} =\mu \frac{B_{0} }{\sqrt{4 \pi} \rho_{0} L_{0}},\hbox{ } \rho^{*}=\frac{\rho }{\rho_{0}} ,\hbox{ }t^{*} = t \frac{ v_{A0}}{L_{0}}, \hbox{ } l^{*}  =                 \frac{l}{L_{0} }, P^{*} = P\frac{ 4 \pi}{B_{0}^{2} }, \hbox{ } P_{c}^{*} = P_{c}\frac{ 4 \pi}{B_{0}^{2} }, \hbox{ } T^{*} = \frac{T}{T_{0}^{A}}, \nonumber \\
       &   \mathbf{v}^{*} = \frac{\mathbf{v}}{v_{A0}}, \hbox{ } \mathbf{B}^{*} =  \mathbf{B}\frac{ \sqrt{4 \pi}}{B_{0}}, 
       \hbox{ } \mathbf{x}^{*}  =  \frac{\mathbf{x}}{L_{0}}, \hbox{ }\mathbf{k}^{*}=  \mathbf{k} L_{0} , \hbox{ }                    \widehat{\mathbf{b}}^{*}=  \widehat{\mathbf{b}}, \nonumber \\
       &   \eta^{*} =  \frac{\eta}{\rho_{0} L_{0} v_{A0}}=\eta_{c}^{*}\left(T^{*}\right)^{5/2},\hbox{with } \eta_{c}^{*}           =\eta_{c}\frac{\left(T_{0}^{A}\right)^{5/2}}{\rho_{0} L_{0} v_{A0}}, \nonumber \\
       &   \kappa^{*} = \kappa \frac{\overline{m}}{\rho_{0} L_{0} v_{A0}} = \kappa_{c}^{*}\left(T^{*}\right)^{5/2},             \hbox{with } \kappa_{c}^{*} = \kappa_{c}\frac{\overline{m}\left(T_{0}^{A}\right)^{5/2}}{\rho_{0} L_{0} v_{A0}}.        \nonumber \\
\end{eqnarray}
%
%
With these new variables, the dimensionless thin flux tube equations have exactly the same form as Eq. \eqref{eqn:TFT_mass}, \eqref{eqn:TFT_mom}, \eqref{eqn:TFT_energ}, and \eqref{eqn:TFT_P0_time}, except that the Boltzmann constant is no longer present. From now on, we will refer as the thin flux tube equations as the dimensionless ones.

%
%
%
%
%
%

\section{Solutions of Reconnection Dynamics}
\label{sec:sims} 

\subsection{Low Plasma-$\beta$ Uniform Background Case}
   \label{sec:Unif_Low_beta}

The results of the last section are quite general. In this section, we will simplify the equations to apply them to the low-$\beta$ solar corona. 
For simplicity, we will also assume uniform background magnetic field. 

\begin{figure}
  \centering
   \includegraphics{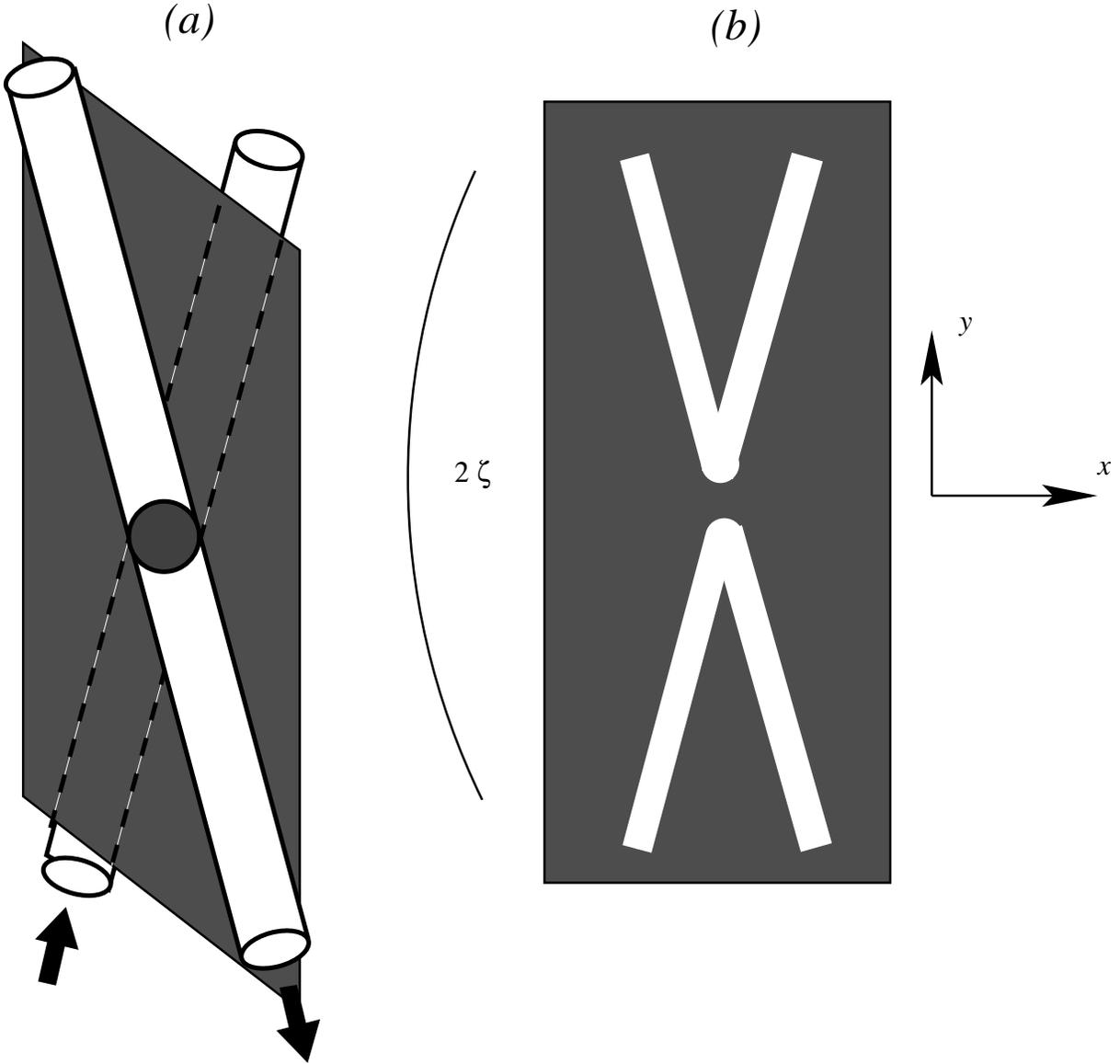}
   \caption{Current sheet cartoon. (a):The two tubes represent the front (solid contour) and back (dashed contour) pre-reconnection tubes. Thick arrows indicate the direction of the magnetic field. The sphere in the middle represents the small reconnection region. The tubes are initially part of the uniform background equilibrium.  (b): Reconnected V-shaped tubes a short time after reconnection. The angle between the initial tubes is $2 \zeta$. Due to magnetic tension they will retract in opposite directions. The end-points of the flux tubes are located near the edges of the current sheet. }
   \label{fig:current_sheet}
\end{figure}

We will consider 
skewed, uniform external magnetic field forming an angle $2 \zeta$ on opposite
sides of a static, infinite current sheet, as shown in Fig. \ref{fig:current_sheet}. The current sheet is assumed to be infinitely thin, represented by the x-y plane. The initial pressure is also assumed to be uniform everywhere. A spatially and temporally localized reconnection event has occurred (by an unspecified physical mechanism), connecting some field lines from one side of the current sheet, to some on the other side. The amount of flux is determined by the size of the reconnection region. This bundle of field lines forms two V-shaped flux tubes (as shown in panel (b) of Fig. \ref{fig:current_sheet}) and magnetic tension at their cusps causes them to retract. These tubes will remain as coherent entities, preserving their magnetic flux $\phi$, while they move. No further reconnection is assumed. 

For this uniform case, the characteristic magnetic field magnitude and density at the reconnection point are the same as everywhere else. Then, $B_{0}=B_{e}$, $\rho_{0} = \rho_{e}$, and $L_{0}$ is some characteristic length. The perpendicular gradient of the external total pressure is zero. 

In the solar corona, the ratio between thermal pressure and magnetic pressure, $\beta = 8\pi P/B^{2}$, is small.  For a plasma with magnetic field $B_{e} =$ \realB, number density $n_{e} =$ \realnumdens, and temperature $T_{e} =$ \realTemp, $\beta$ is approximately equal to \realbeta. Consequently, we will neglect the thermal pressure with respect to the magnetic pressure. The initially small internal pressure might increase due to compressions along the tube, but we will consider it small compared to the internal magnetic pressure at all times. In the parallel direction, there is no force related to the magnetic field, therefore the parallel derivative of the thermal pressure cannot be neglected. 

With the above considerations, and making explicit the temperature dependence of the transport coefficients, we can re-write the constraints and thin flux tube equations in the following way
%
%
%
\begin{eqnarray}
    \label{eqn:TFT_low_beta_B}
       B  &= & B_{e}, \\
    \label{eqn:TFT_low_beta_mass}
       \frac{D\rho}{Dt} & = & -\rho \widehat{\mathbf{b}} \cdot \frac{\partial \mathbf{v}}{\partial l}, \\
    \label{eqn:TFT_unif_cons_mom}
       \frac{D\mathbf{v}}{Dt}  &= & \frac{\partial}{\partial \mu} \left\{\left[ \frac{B_{e}}{4 \pi} - \frac{P}{B_{e}}                + \frac{\eta_{c}T^{5/2}}{B_{e}} \left( \widehat{\mathbf{b}} \cdot \frac{\partial \mathbf{v}}{\partial l}                      \right ) \right] \widehat{\mathbf{b}} \right\} + \frac{P}{\rho} \mathbf{k} \\
       &= &   
       \frac{1}{\rho} \left\{ 
                       \frac{B^{2}_{e}}{4 \pi} \mathbf{k} 
                      -\widehat{\mathbf{b}} \frac{\partial P}{\partial l} - P \mathbf{k}
                      + \frac{\partial}{\partial l} \left[ \widehat{\mathbf{b}} \eta_{c}T^{5/2}                                             \left(\widehat{\mathbf{b}}\cdot \frac{\partial\mathbf{v}}{\partial l}\right) \right]
                      \right\} + \frac{P}{\rho}\mathbf{k}, \nonumber \\
    \label{eqn:TFT_unif_cons_ene}
       \frac{D \varepsilon_{T}}{Dt} &= & \frac{\partial}{\partial \mu} \left\{\left[\frac{B_{e}}{4 \pi} -                      \frac{P}{B_{e}} + \frac{\eta_{c}T^{5/2}}{B_{e}}\left( \widehat{\mathbf{b}} \cdot \frac{\partial               \mathbf{v}}{\partial l}\right )\right ] \left( \widehat{\mathbf{b}} \cdot \mathbf{v}\right ) + 
       \frac{\kappa_{c}T^{5/2}}{B_{e}}\frac{\partial T}{\partial l}\right\} + \frac{P}{\rho} \mathbf{k} \cdot \mathbf{v}, 
\end{eqnarray}
%
%
where
%
%
\begin{eqnarray}
    \label{eqn:TFT_spec_ene_I}
       \varepsilon_{T} = \varepsilon_{I} + \varepsilon_{K} + \varepsilon_{M},
\end{eqnarray}
with specific internal, kinetic, and magnetic energies
\begin{eqnarray}
    \label{eqn:TFT_spec_ene_II}
       \varepsilon_{I} = \frac{P}{(\gamma -1) \rho},\hbox{ } 
       \varepsilon_{K} = \frac{|\mathbf{v}|^{2}}{2},\hbox{ } 
       \varepsilon_{M} = \frac{B_{e}^{2}}{4 \pi \rho}.
\end{eqnarray}
%
%
%
The specific magnetic energy of the tube doubles because there is work done by the background magnetic field expanding into volume vacated by motion of the tube. We have assumed that the external field returns to its original value after the tube has passed. In  more realistic scenarios this value may be slightly different. 

Conservation of energy improves stability of any numerical solution. Toward this end, we will neglect the small last term in Eq. \eqref{eqn:TFT_unif_cons_ene} (proportional to $\beta$), that is not part of the perfect derivative in the integrated mass. With this simplification, Eq. \eqref{eqn:TFT_P0_time} remains unchanged, and the change of the tube's total energy per flux is equal to
%
%
%
\begin{eqnarray}
   \label{eqn:TFT_cons_Energy} 
   \frac{D }{Dt} \left( \frac{E_{T} }{\Phi}\right ) &=& \left. \left\{ \left[\frac{B_{e}}{4 \pi} - \frac{P}{B_{e}} + \frac{\eta_{c}T^{5/2}}{B_{e}}\left      
   (\widehat{\mathbf{b}} \cdot \frac{\partial \mathbf{v}}{\partial l}\right )\right ] \left( \widehat{\mathbf{b}} \cdot \mathbf{v}\right ) + 
   \frac{\kappa_{c}T^{5/2}}{B_{e}}\frac{\partial T}{\partial l} \right \} \right |^{\mu_{0}}_{\mu_{L}}, 
\end{eqnarray}
%
%
%
where ${\mu_{0}}$ and ${\mu_{L}}$ represent the end points of the tube. If the end points are fixed, and there is no temperature gradient there, the total energy of the tube is conserved.  The neglected term came from the change in kinetic energy, calculated from Eq. \eqref{eqn:TFT_unif_cons_mom}, and to be consistent, the correspondent perpendicular term will be neglected in the momentum equation, as well. Then, the change in total momentum per flux of the tube is equal to 
%
%
\begin{eqnarray}
  \label{eqn:TFT_cons_momentum}
   \frac{D }{Dt} \left( \frac{\mathbf{P}_{T} }{\Phi}\right ) &=& \left. \left\{\left[\frac{B_{e}}{4 \pi} - \frac{P}{B_{e}}   
   + \frac{\eta_{c}T^{5/2}}{B_{e}}\left( \widehat{\mathbf{b}} \cdot \frac{\partial \mathbf{v}}{\partial l}\right )
   \right ]  \widehat{\mathbf{b}} \right \} \right |^{\mu_{0}}_{\mu_{L}}. 
\end{eqnarray}
%
%
With the neglect of the proportional to $\beta$ small term, and the fact that $P_{e}$ is uniform, the ideal part of Eq. \eqref{eqn:TFT_unif_cons_mom} reduces to the thin flux tube equation presented by \citet{Longcope_2009}. Therefore, gas-dynamics shocks and rotational discontinuities are expected along the reconnected flux tubes.

The volumetric heating rate when the magnetic field magnitude is uniform becomes
%
%
\begin{eqnarray}
   \label{eqn:heating_uniform}
      \dot{Q} & = &  \eta_{c} T^{\frac{5}{2}} \left(\widehat{\mathbf{b}} \cdot \frac{\partial \mathbf{v}}{\partial l} 
      \right)^{2} +  \frac{2}{7} \kappa_{c} \frac{\partial^{2}T^{\frac{7}{2}}}{\partial l^{2}}.
\end{eqnarray}
%
In the above equation, viscosity always contributes to heat the plasma. Depending on the curvature of temperature to the $7/2$ power, and the value of the Prandtl number, the thermal conduction term may contribute with plasma cooling or heating.  

%
%
%
\subsection{Simulations}
   \label{sec:Simulations}

We developed a computer program, called DEFT (Dynamical Evolution of Flux Tubes), that solves the thin flux tube equations for the retraction of the two reconnected thin flux tubes described in last section. The program takes into account the effect of the transport coefficients, including their strong dependence on temperature. It also can perform simulations using real coronal dimensionless parameters, as well as implement a strong anisotropy in the transport coefficients (heat conduction and momentum transport are allowed only in the direction parallel to the magnetic field). This is far more difficult to achieve in fully 3D MHD programs. The DEFT program can be implemented for different background conditions as well, but we will restrict ourselves in this paper to the low $\beta$ uniform background case from last section. In this case, the two tubes moving in opposite directions are symmetric.  In cases with non-uniform background that might not be so. 
We use the conservative form of the thin flux tube equations introduced in the last section. 

The DEFT program uses a staggered mesh where each tube piece is represented by grid points at its ends, and by its mass per flux. The code implements a Lagrangian approach where each tube piece is followed, guaranteeing mass conservation. 
Due to mirror symmetry, only half of the tube is simulated, the other half is a mirror image in the x-direction.

The initial state corresponds to two opposite V-shaped flux tubes, connected at the reconnection region, as shown in Fig. \ref{fig:current_sheet}. The initial angle between the magnetic field lines at each side of the current sheet is equal to $2 \zeta$. Each tube is divided in $N$ segments separated by $N+1$ tube-point positions. The initial tubes are clearly out of equilibrium due to the sharp angle between the field lines. To avoid introducing length scales at the limit of resolution, the central parts of the initial tubes are smoothed. The end-points of the tube are assumed to be fixed, and to have no temperature gradient with the external boundaries (this last condition ensures no heat transfer from the end-points).

Since the reconnected tubes originated as background plasma they are initialized with uniform density, $\rho = \rho_{e}$, and thermal pressure $P = P_{e}$. The mass per flux of each piece is equal to $\delta m / \Phi = \rho_{e} \delta l_{e} /B_{e}$, where $\delta l_{e}$ is the segment's initial length. The magnetic field magnitude, $B_{e}$, is constant everywhere throughout the entire simulation.
The initial velocities are zero (equilibrium state), even for the central bend (we are not interested in the details of the initial transient evolution). Viscosity and thermal conductivity are calculated using Eq. \eqref{eqn:transp_coeff_eta} and \eqref{eqn:transp_coeff_kappa}, where $T = P/\rho$ (unitless quantities). The unit parallel vectors are calculated at segment's centers from definition \eqref{eqn:bvect}, and curvature vectors are calculated at grid points from definition \eqref{eqn:kvect}. Arc-length derivatives are computed using centered differences.

At every time, the $N+1$ positions and velocities, as well as the $N$ pressure pre-factors $P_{c}$, are advanced using the current state variables in Eq. \eqref{eqn:TFT_unif_cons_mom} and \eqref{eqn:TFT_P0_time}, following a simple Eularian approach, for example $ \mathbf{v}(t + \delta t) = \mathbf{v}(t) + (D/Dt)\mathbf{v}(t) \delta t$. The time interval is chosen to satisfy the Courant-Friedrichs-Lewy 
condition \citep{CFL_1967}.
 
All terms in Eq. \eqref{eqn:TFT_unif_cons_mom} are evaluated at grid points, while terms in 
Eq. \eqref{eqn:TFT_P0_time} are evaluated within segments. The state variables and the unit tangent vector $\widehat{\mathbf{b}}$ are defined inside the segments. They are averaged between neighbors when needed at grid points. A similar approach is used when vectors defined at tube points are needed at segment centers. 
From the advanced positions, velocities, and pressure pre-factors $P_{c}$, it is possible to define all the advanced state variables. The density of each segment at a given time is calculated as $\rho = (\delta m/ \Phi) (B_{e} / \delta l) $, where $\delta l$ is the updated length of the segment. Pressure follows from Eq. \eqref{eqn:TFT_P0} and \eqref{eqn:TFT_P0_time}, and the process is repeated until the end of the simulation.

As the two tubes retract they move in opposite directions, but due to the symmetry of the background configuration their dynamics are identical. Therefore, we present results only for the downward moving one. The simulated equations are dimensionless, but for concreteness the results are re-dimensionalized using characteristic magnetic field magnitude, density and length.  For this illustrative purpose we choose $B_{0} =$ \realB, density $\rho_{0} = n_{0} \overline{m}$ with electron number density $n_{0} =$ \realnumdens, and $\overline{m}$ equal to half the mass of the proton. The initial angle between the pre-reconnection field lines is $2 \zeta=$ \doublezeta. The associated Alfv\'{e}n speed and temperature are $v_{A0} \simeq$ \realvao, and $T_{0}^{A} \simeq$ \realTao. We will consider a plasma at $T_{e}$ = \realTemp, and the plasma-$\beta$ in this case is equal to $\beta = {2 T_{e}}/{T_{0}^{A}}\simeq$ \realbeta. For a typical length of $L_{0} =$ \realL, the Reynolds number is $R_{\eta}\simeq$ \realReynolds. An Alfv\'{e}n time (time to move one characteristic length at the Alfv\'{e}n speed) would be approximately \realalfvtime. As mentioned before, the Prandtl number of the solar corona is typically equal to \PrandtlSim, and we will use this value in our simulations. The total number of segments in the tube is \numsegments. 

The simulation is carried out for approximately \fintime, to allow for the full development of shocks (discussed in detail in the next section). The initial length of the tube was chosen to be unrealistically long in order to achieve the steady state for the shocks. As we will see in the next section, the thickness of the shocks, when fully developed, is of the order of the length of an entire coronal loop. The fluid inside the tube evolves towards the steady state, which it probably would not attain within a realistic flare time. 

\begin{figure}
   \centering
   \includegraphics{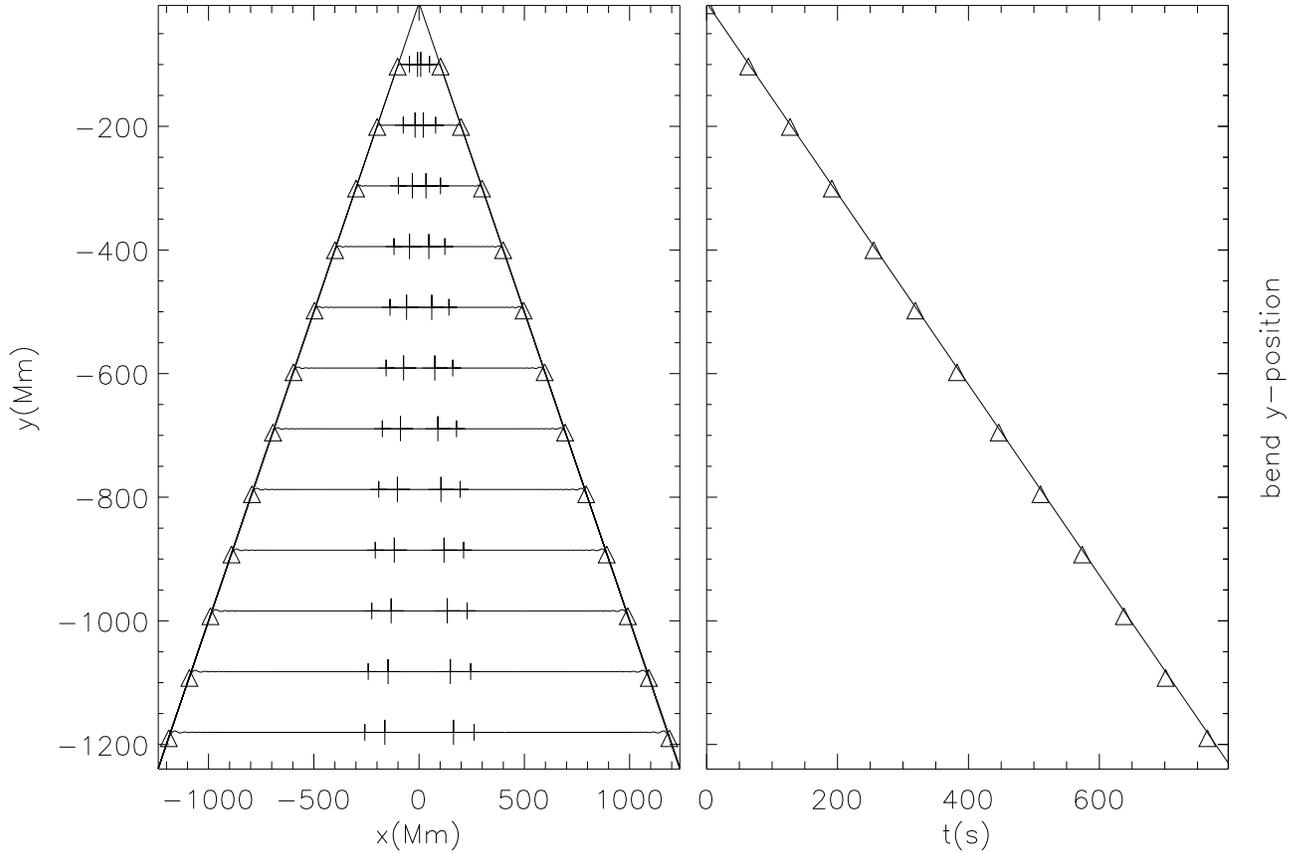}
   \caption{Left panel: Simulation results for the temporal evolution of a downward moving reconnected tube. Each horizontal segment corresponds to a different time. Vertical ticks account for the thickness of the two outward moving gas-dynamics shocks. The width of the shock evolves up to a steady value. Triangles indicate bend positions. Right panel: right bend y-position as function of time (same times chosen for the left panel). The solid line correspond to a line with a slope equal to the y-projection of the Alfv\'{e}n speed at the reconnection point, $v_{Ae}$. Both panels share the same y-axis scale}
   \label{fig:tube_evolution}
\end{figure}

The left panel of Fig. \ref{fig:tube_evolution} shows the downward moving tube at different times.  The tube is seen to be composed of three main segments, one horizontal segment moving downward at the y-projection of the Alfv\'{e}n speed, and two legs from the unperturbed parts of the initial tube. The two corners where these three pieces come together move at the Alfv\'{e}n speed along the legs of the initial tube and will be called {\em bends} (triangle shapes in the Figure). In the right panel, the y-position of the right bend is plotted as function of time for the same times as in the left panel. The solid line correspond to a line with a slope equal to the y-projection of the Alfv\'{e}n speed at the reconnection point ($v_{Ae}$). 

\begin{figure}
  \centering
   \includegraphics{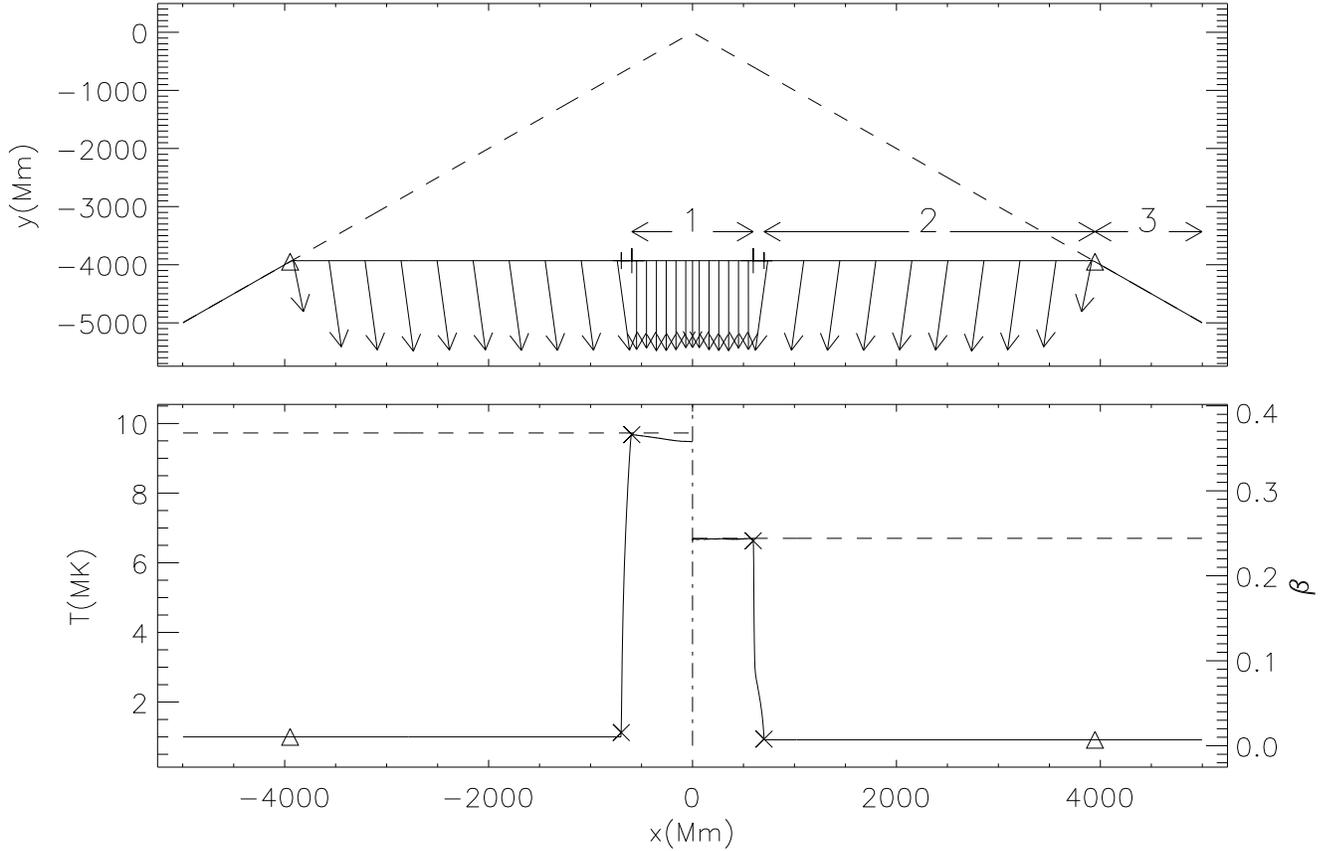}
   \caption{Top panel: Dashed line represents initial tube. Solid line represents the tube at time equal to \timeSim \ (vertical scale is compressed). Horizontal arrows delimit regions 1, 2 and 3 (post-shock plasma, inflow region, and unperturbed leg of the tube, respectively). Arrows show velocity directions in the tube (not to scale). Plasma is deflected in the bisector direction at the bends. Large and small vertical ticks indicate the beginning and end of the width of the shock, respectively. Bottom panel: the left side depicts temperature profile in half of the tube, and the right side, the plasma-beta profile for the same time as in the top panel. Dashed lines show the theoretical steady-state Rankine-Hugoniot post-shock values. Crosses are the same as vertical ticks in the top panel. In both panels triangles indicate the position of the bends. There is no change in the state variables at the bends (Alfv\'{e}n waves). The tube has mirror symmetry in the x-direction. Both panels share the same x-axis scale}
   \label{fig:profile_T_Beta}
\end{figure}

The magnitude of the magnetic field remains constant along the tube due to constraint \eqref{eqn:TFT_low_beta_B}, therefore the area of the tube also remains constant (Eq. [\ref{eqn:flux_tft}]). At the bends, the initially stationary plasma is deflected along the bisector of the angle between the two adjacent straight segments (average direction of the curvature force), as shown in Fig. \ref{fig:profile_T_Beta}. In this Figure, a time approximately equal to \timeSim\ was chosen to show the shape and internal structure of the tube long after reconnection. 

The velocity of the deflected plasma (region 2 in Fig. \ref{fig:profile_T_Beta}) is \citep[see][]{Longcope_2009}
%
%
\begin{eqnarray}
    \label{eqn:inflow_speed}
       \mathbf{v}_{2}  = - 2 v_{Ae} \sin^{2}\left( \frac{\zeta }{2} \right) \widehat{\mathbf{x}} - v_{Ae} \sin (\zeta)  \widehat{\mathbf{y}}.
\end{eqnarray}
%
%
This velocity is depicted in the top panel of the mentioned Figure (the left side is a mirror image). 

The parallel component ($\widehat{\mathbf{x}}$ direction) of this velocity is proportional to the Alfv\'{e}n speed, and is therefore generally supersonic ($v_{Ae} = c_{se} \sqrt{2/\gamma \beta}$). The tube gets shorter as it moves, creating parallel supersonic flows at the bends. These supersonic flows are directed toward the center of the tube where they collide. The small, but non-zero thermal pressure there prevents plasma from piling up, and stops it in the parallel direction. The two colliding flows generate two gas-dynamic shocks (the magnetic field remains constant across these shocks) that move outward from the center of the tube. In the top panel of Fig. \ref{fig:profile_T_Beta}, velocity discontinuities in the parallel direction between region 1 (post-shock region) and 2 (inflow region), are evident. The vertical ticks indicate the beginning and end positions of these shocks. In the bottom panel of the Figure, the left side depicts the temperature profile along the tube, and the right side, the plasma-$\beta$ profile, both for t = \timeSim. We see that temperature rises almost one order of magnitude from the pre-shock value, and the plasma-$\beta$ more than $30$ times its initial value, and neither changes at the bends since these corners are Alfv\'{e}n waves.

\begin{figure}
  \centering
   \includegraphics{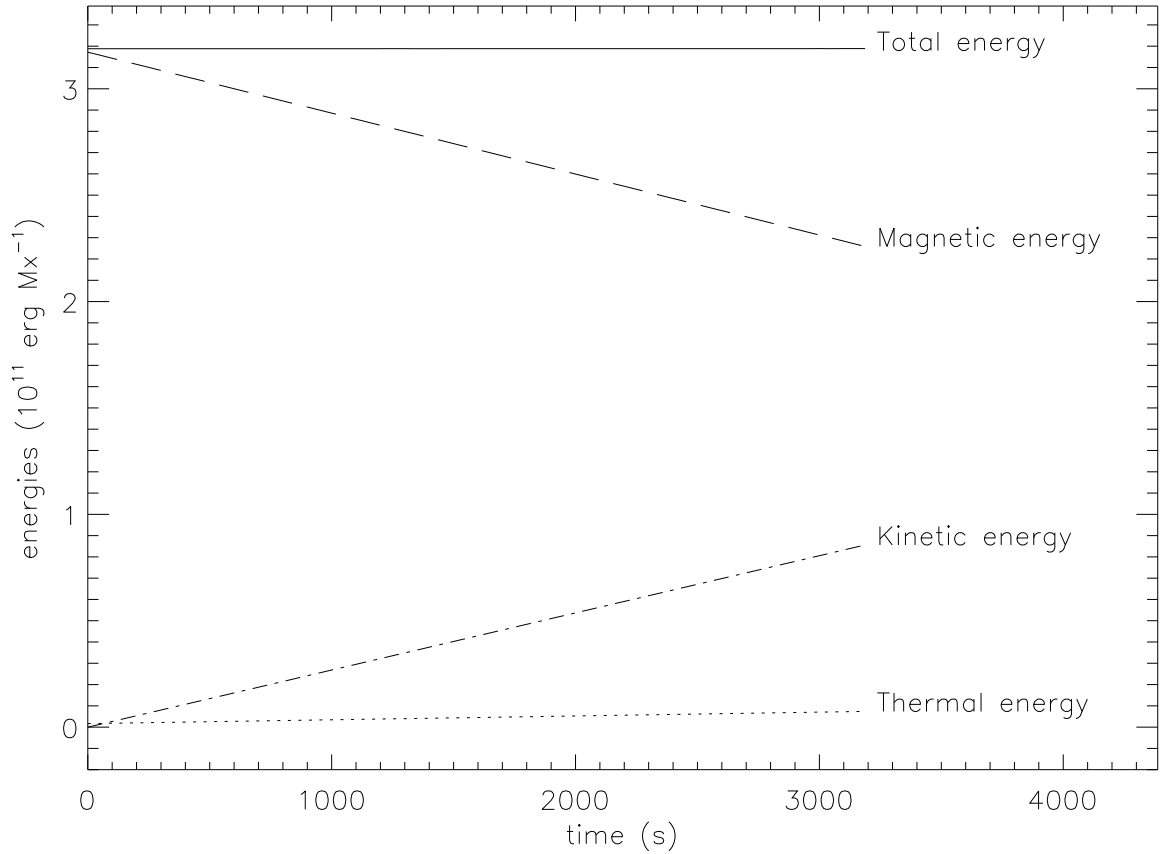}
   \caption{ Energies' temporal evolution for the simulation presented in Section \ref{sec:Simulations}. Each line represents the temporal evolution of the corresponding specific energy integrated over the entire tube. The percentage of magnetic energy converted to thermal is small (less than 10 \%). The rest is converted to kinetic energy. }
   \label{fig:energy_convertion}
\end{figure}

From Eq. \eqref{eqn:TFT_cons_Energy} and the boundary conditions at end-points of the tube, we see that the total energy of the tube is conserved (as shown in Fig. \ref{fig:energy_convertion}). Even though, temperature changes significantly at the shock, only a small fraction (less than 10 \%) of the available magnetic energy is converted to thermal energy (Fig. \ref{fig:energy_convertion}). The magnetic energy of the tube decreases with time as the tube shortens (the magnitude of the field remains constant in the tube, but its length decreases), and is mostly converted to kinetic energy at the bends.

%
%
\section{Gas-Dynamic Shock Internal Structure}
   \label{sec:Inner_Structure}

Transport coefficients (viscosity and thermal conduction) and their temperature dependence determine the inner structure of the gas-dynamic shocks. 
As the flux tubes retract, they follow a time dependent evolution until they reach a steady-state characterized by a transition between upstream and downstream states dictated by the Rankine-Hugoniot conditions \citep{Rankine_1870,Hugoniot_1887}. The diffusive terms smooth out the transitions but do not alter the upstream or downstream values provided these regions are asymptotically uniform (${\partial \mathbf{v}}/{\partial l}\to0$ and ${\partial T}/{\partial l}\to0$).  These asymptotic values are governed by conservations laws observed by the diffusion.

Gas dynamics shocks in reconnected thin flux tubes were introduced by \citet{Longcope_2009}, in the case of ideal thin flux tube equations. It was shown there that only two kind of discontinuities are supported by thin, reconnected thin tubes embedded in a low plasma-$\beta$ uniform background: one that corresponds to a bend moving at the Alfv\'{e}n speed, where temperature, density and pressure are continuous, and a second one that corresponds to a gas-dynamic shock within a straight section of the tube. Both of these discontinuities were present in our simulation. In this section, we will focus on the later one, where heating and cooling occur (the heating at the bends is negligible). 

To determine the effect of diffusive terms in the inner structure of the shocks, we follow an approach similar to the one used by several authors \citep{Thomas_1944,Grad_1952,Gilbarg_1953,Kennel_1988}. We consider a straight section 
($\partial \widehat{\mathbf{b}}/ \partial l = 0$) and further assume the presence of a 
transition (shock) moving to the right at constant speed $v_{s}$. The inflow region will be denoted by a ``2'' subscript, and the post-shock region by a ``1'' subscript. In our reconnection scenario, the inflow region corresponds to the plasma moving supersonically after being accelerated  by the bend (region 2 in Fig. \ref{fig:profile_T_Beta}), and the post-shock region corresponds to the center of the tube (region 1 in the same Figure).  

\subsection{Steady-state Solution}

The Mach number (the ratio between the fluid velocity and the local sound speed) in the inflow region is determined by the initial plasma-$\beta$ and the initial angle between the field lines \citep{Longcope_2009}, namely
%
%
\begin{eqnarray}
   \label{eqn:Mach_num} 
      M_{2} = \sqrt{ \frac{8 }{\gamma \beta_{e}} } \sin^{2} \left( \frac{\zeta }{2} \right),
\end{eqnarray}
%
%
and the Mach number in the reference frame of the shock is 
%
\begin{eqnarray}
   \label{eqn:Mach_num_shock} 
      M_{2,s} = M_{2} + \frac{v_{s}}{c_{s,2}}=\sqrt{\frac{u_{2}^{2}}{\gamma T_{2}}}, 
\end{eqnarray}
%
%
where $c_{s,2}$ is the sound speed in region 2, and $u_{2}$ and $T_{2}$ are the pre-shock velocity in the shock reference frame, and temperature, respectively. For our simulation in the last section, the Mach numbers were $M_{2} \simeq$ \MachSim, and $M_{2,s} \simeq$ \MachSimShock.

In the reference frame of the shock, the inflow velocity corresponds to plasma moving at $u_{2} = v_{x,2} - v_{s}$ ($< 0$), and the post-shock flow is $u_{1} = -v_{s}$ (plasma is stopped in the parallel direction). The ratio between the shock speed, and the inflow speed magnitude, $|v_{x,2}|$, is given by
%
\begin{eqnarray}
   \label{eqn:shock_vs} 
       \frac{v_{s}}{|v_{x,2}|}  = \sqrt{ \frac{1}{M_{2}^{2}} + \frac{(\gamma + 1)^{2}}{16} } - \frac{3-\gamma}{4} .
\end{eqnarray}
%
%
If constant states are assumed, $\partial / \partial t = 0$ for all the quantities calculated in the moving frame of this discontinuity. In this reference frame, the steady-state, conservative, unitless TFT equations for uniform background case and low beta can be written as 
%
%
\begin{eqnarray}
    \label{eqn:Int_Str_mass}
       \rho_{1} u_{1} & = & \rho_{2} u_{2} = \rho u , \\
    \label{eqn:Int_Str_mom}
         \rho_{1} u^{2}_{1} + P_{1} & = &  \rho_{2} u^{2}_{2} + P_{2}    = \rho u^{2} + P - \eta \frac{\partial u}{\partial l}, \\
    \label{eqn:Int_Str_ene}
         \rho_{1} u^{3}_{1} + u_{1} P_{1} \frac{2 \gamma}{\gamma-1} & = & \rho_{2} u^{3}_{2} + u_{2} P_{2} \frac{2 \gamma}{\gamma-1} =  \nonumber \\ 
        & = &  \rho u^{3} + u P \frac{2 \gamma}{\gamma-1} - 2 u \eta \frac{\partial u}{\partial l} - 2 \kappa \frac{\partial T}{\partial l},
\end{eqnarray}
%
%
%
The second equalities in each of the above equations refer to the state variables inside the shock. The usual Rankine-Hugoniot conditions can be found from the first equalities in each of the above equations. 

\begin{figure}
  \centering
   \includegraphics{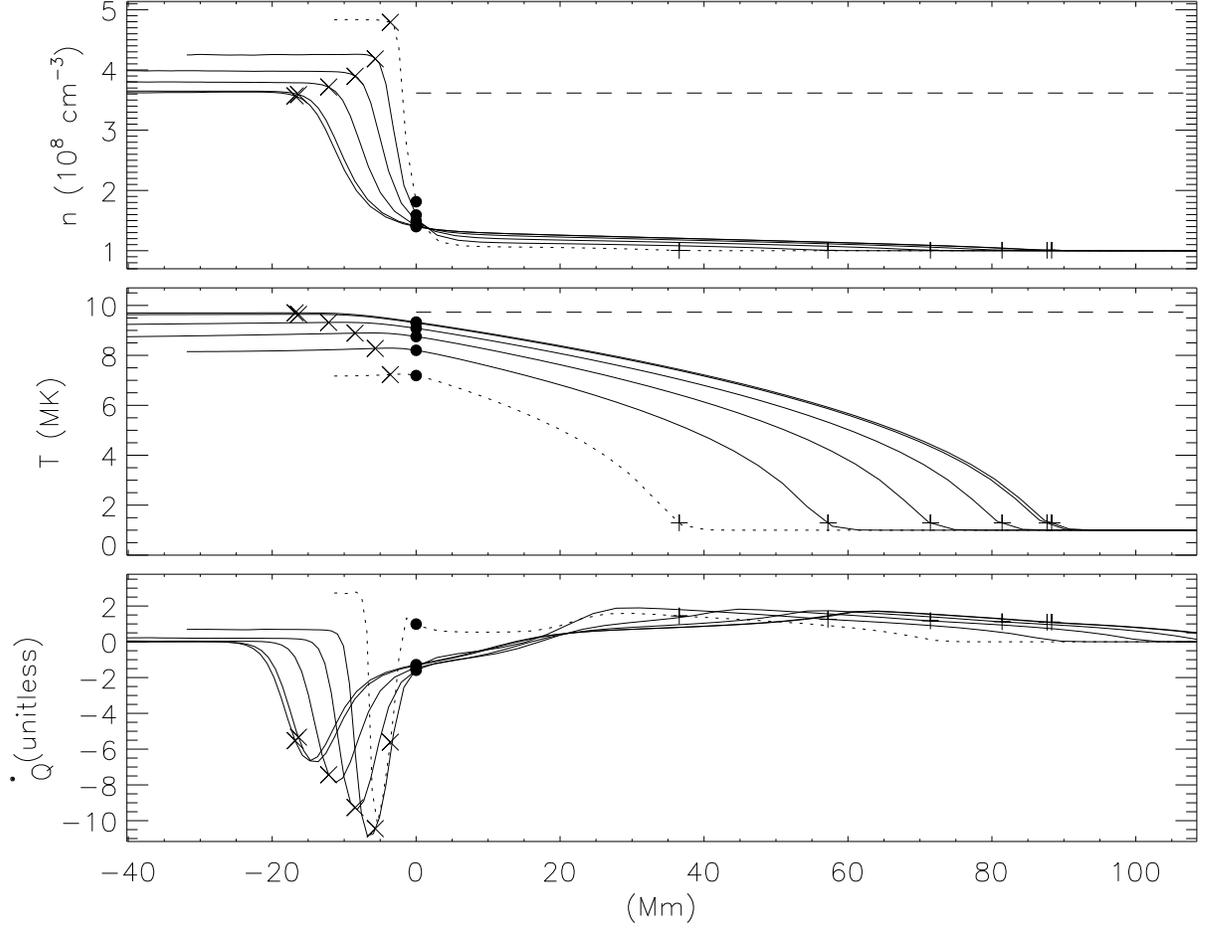}
   \caption{Top panel: Electron number density evolution in the shock frame for times $t = 63$ s, $159$ s, $319$ s, $638$ s, $1913$ s, and $3156$ s. Middle panel: Temperature evolution in the shock frame for the same times as in the top panel. Bottom panel: Volumetric heating rate in the shock frame for the same times as in the top panel. In these three panels, plus signs indicate the position of the heat front, crosses indicate the position of the isothermal sub-shock, and dark dots indicate the position of the transition between the isothermal sub-shock and the thermal front. The dotted lines indicate the earliest time. Horizontal dashed lines are the post-shock Rankine-Hugoniot correspondent values.}
   \label{fig:shock_heating_cooling}
\end{figure}

It is convenient for our analysis of the internal structure of the shock, to define a characteristic velocity $u^{*} = \frac{1}{2} \left(u_{2} + {T_{2}}/{u_{2}} \right)= ({u_{2}}/{2})\left( 1 + 1/{\gamma  M_{2,s}^{2} } \right )$. With this definition, the  Rankine-Hugoniot conditions can be written as 
%
\begin{eqnarray}
   \label{eqn:jump_dens} 
      \frac{\rho_{2}}{\rho_{1}} & = &  \frac{u_{1}}{u_{2}} = \frac{(4 \gamma u^{*}/\gamma+1) - u_{2}}{u_{2}}, \\
   \label{eqn:jump_T}
      \frac{T_{1}}{T_{2}} & = & \frac{u_{1} (2u^{*}- u_{1})}{T_{2}}.
\end{eqnarray}
%
%
The plasma-$\beta$ ratio can be calculated as ${\beta_{1}}/{\beta_{2}} = 
({\rho_{1}}/{\rho_{2}})({T_{1}}/{T_{2}})$. In Fig. \ref{fig:profile_T_Beta}, \ref{fig:shoulder_zoom}, and \ref{fig:shock_heating_cooling}, horizontal dashed lines show the theoretical steady-state jump values for our simulation. Even-though the post-shock plasma-$\beta$ is more than $30$ times the initial value, it is still below unity. 

The second equalities in Eq. \eqref{eqn:Int_Str_mom} and \eqref{eqn:Int_Str_ene}, are differential equations for the shock's internal state variables. Their boundary conditions are given by the steady-state Rankine-Hugoniot conditions. Combining these two equations we obtain
%
%
\begin{eqnarray}
   \label{eqn:inner_prandtl} 
      \frac{2(\gamma-1)}{P_{r} \hbox{ } u} \frac{\partial T}{\partial u} & = & \frac{2 T - u^{2} (\gamma-1) + 4 (\gamma-1) u^{*}u + u_{2}^{2} (\gamma+1) - 4 \gamma u_{2} u^{*}}{u^{2} - 2 u u^{*} + T} = \frac{h(T,u)}{g(T,u)},
\end{eqnarray}
%
%
where we have defined functions $h(T,u)$ and $g(T,u)$ as the numerator and denominator of the above differential equation. $P_{r}$ indicates the Prandtl number. This equation is valid for any temperature dependence of the transport coefficients, as long as both are the same; and can be numerically integrated for different Prandtl numbers and shock strengths. Numerical solutions of this equation in conjunction with Eq. \eqref{eqn:Int_Str_mom} or \eqref{eqn:Int_Str_ene} provide a method to determine the thickness of the shock.

\begin{figure}
  \centering
   \includegraphics{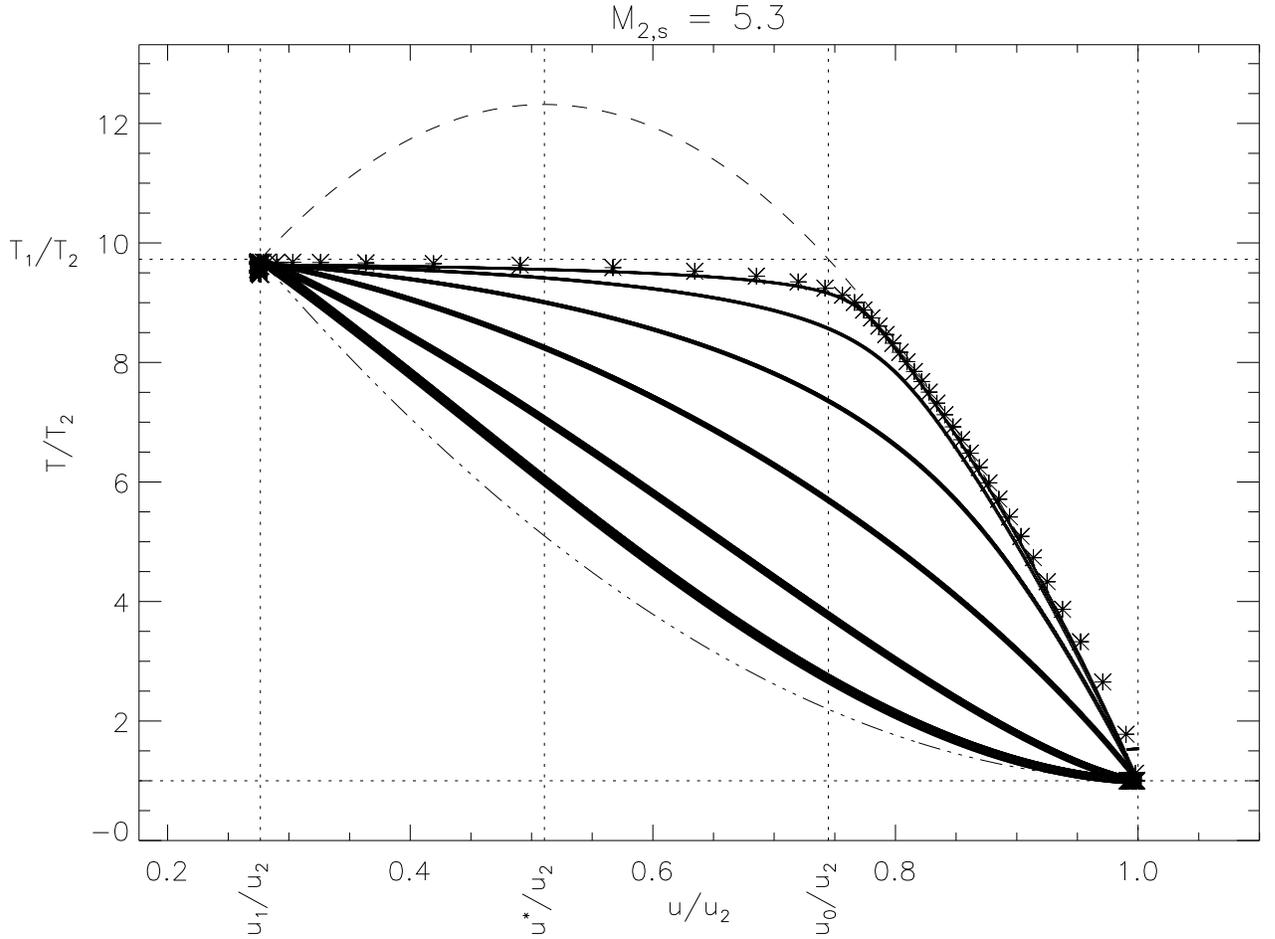}
   \caption{ Fixed Point Diagram for super-critical Mach number equal to \MachSimShock\ (the same one than for simulation in Section \ref{sec:Simulations}). Temperature (ordinate) and velocity (abscissa) are scaled to the pre-shock values. For this case, $|u^{*}|>|u_{1}|$. Top and bottom horizontal dotted lines represent post-shock and pre-shock temperatures, respectively. The leftmost vertical dotted line corresponds to the post-shock speed, and the rightmost dotted line the pre-shock speed. The dashed line represents the curve $g(T,u)=0$, and the mixed dotted and dashed line depicts the $h(T,u) =0$ curve (the latter is the limit solution for an infinite Prandtl number). Each solid line represents a numerical solution for equation \eqref{eqn:inner_prandtl} for different Prandtl numbers (increasing thickness of the line corresponds to increasing value of  the Prandtl number). The shown Prandtl numbers are \PrandtlSuperCrit. The second vertical dotted line from the left shows the characteristic speed $u^{*}$ (maximum of the curve $g(T,u)=0$), and the second dotted line from the right shows velocity $u_{0}$ (separation between sub-shock and heat front for $P_{r} = 0$). Asterisks corresponds to grid points inside the shock for the simulation presented in Section \ref{sec:Simulations} ($P_{r}=$ \PrandtlSim) for time = \timeSim\ (the same time chosen for Fig. \ref{fig:profile_T_Beta}). The internal structure of the shock is well resolved by the DEFT program (there are several grid points in each shock section).}
   \label{fig:fixed_point}
\end{figure}

In Fig. \ref{fig:fixed_point}, numerical solutions to the above equation are plotted for the same Mach number as in our simulation from last section (temperature and velocity are normalized to the pre-shock values). The pre-shock and post-shock values are fixed by the strength of the shock, independently of the transport coefficients, therefore we will call this kind of diagrams {\em fixed point diagrams} (they correspond to a velocity contrast-temperature plane described by  \citet{Kennel_1988}, or Navier-Stokes direction fields by \citet{Grad_1952}). Each solid line corresponds to a numerical solution to the above equation for a given Prandtl number (increasing thickness of the line represents increasing value in the Prandtl number). In this Figure, the mixed dotted and dashed line shows the curve $h(T,u) = 0$ that corresponds to the limit of infinite Prandtl number $P_{r}$. As the Prandtl number increases from low values, the solutions (solid lines) approach this limit case. The dashed line represents the $g(T,u)=0$ curve, and asterisks indicate grid points (inside the shock region) from our simulation ($P_{r} = $ \PrandtlSim) for the same time chosen for Fig. \ref{fig:profile_T_Beta}. Simulations coincide with the steady-state theoretical predictions. There are several grid points in each sub-region of the shock, which shows that the DEFT program can resolve the internal structure of the shock.

Velocity $u^{*}$ corresponds to the maximum value of the curve $g(T,u)=0$ in the $T$-$u$ space, as shown in Fig. \ref{fig:fixed_point}. When the magnitude of this speed is smaller than the magnitude of the post-shock speed $u_{1}$, the temperature increases monotonically from $T_{2}$ to $T_{1}$, without presenting an isothermal region, for all Prandtl numbers \citep[\S 88]{Landau_1959}. This occurs in weak shocks, called sub-critical, which Mach numbers $M_{2,s}$ smaller than a critical Mach number $M_{2,cr}=\sqrt{({3\gamma-1})/{(3-\gamma)\gamma}}$. 

On the other hand, if the shock is strong, the fixed point diagram looks like the one depicted in Fig. \ref{fig:fixed_point}. In this case, if the Prandtl number is small (as in the solar corona and our simulation), the solution is composed of two regions \citep[\S 88]{Landau_1959}. In the first region, called the heat front or thermal conduction front, temperature is increased approximately along the $g(T,u) = 0$ curve, from $T_{2}$ up to approximately the steady-state value $T_{1}$. The second region, called isothermal sub-shock or isothermal discontinuity (or jump), consists of an almost isothermal (${\partial T}/{\partial u}$ = 0) change of velocity from $u_{0} = {T_{1}}/{u_{1}}$ to the steady-state value $u_{1}$. This last region corresponds to an isothermal compression of the plasma; most of the compression across the shock occurs in this region (see top panel of Fig. \ref{fig:shock_heating_cooling}). The heat front increases the temperature of the plasma without changing the density significantly. The compression ratio between the initial density $\rho_{2}$ and the density at the transition between the sub-shock and the heat front $\rho_{0}$ is bounded between the values $({\gamma+1})/{2}$ (limit for a infinitely strong shock) and $({3 \gamma-1})/({\gamma +1})$ (at the critical Mach number), for a Prandtl number equal to zero. For a gas constant equal to $\gamma = \frac{5}{3}$, this corresponds to a density ratio between $1.33$ and $1.5$.

\subsection{Physical size}
 
The thickness of the two internal shock regions, sub-shock and heat front, is determined by Eq. (\ref{eqn:Int_Str_mom}) and (\ref{eqn:Int_Str_ene}), respectively. For the case of strong shocks at low Prandtl number, the thickness of the isothermal sub-shock is determined mostly by viscosity, and the heat front length by the thermal conduction, as shown below. To calculate these thicknesses analytically, we approximate our solution by the case where $P_{r}$ is zero. In Fig. \ref{fig:tube_evolution} and \ref{fig:profile_T_Beta}, large and small vertical ticks indicate the beginning and end of the thickness of the shock that correspond to the width of the sub-shock in addition to the width of the thermal front. 

\begin{figure}
  \centering
   \includegraphics{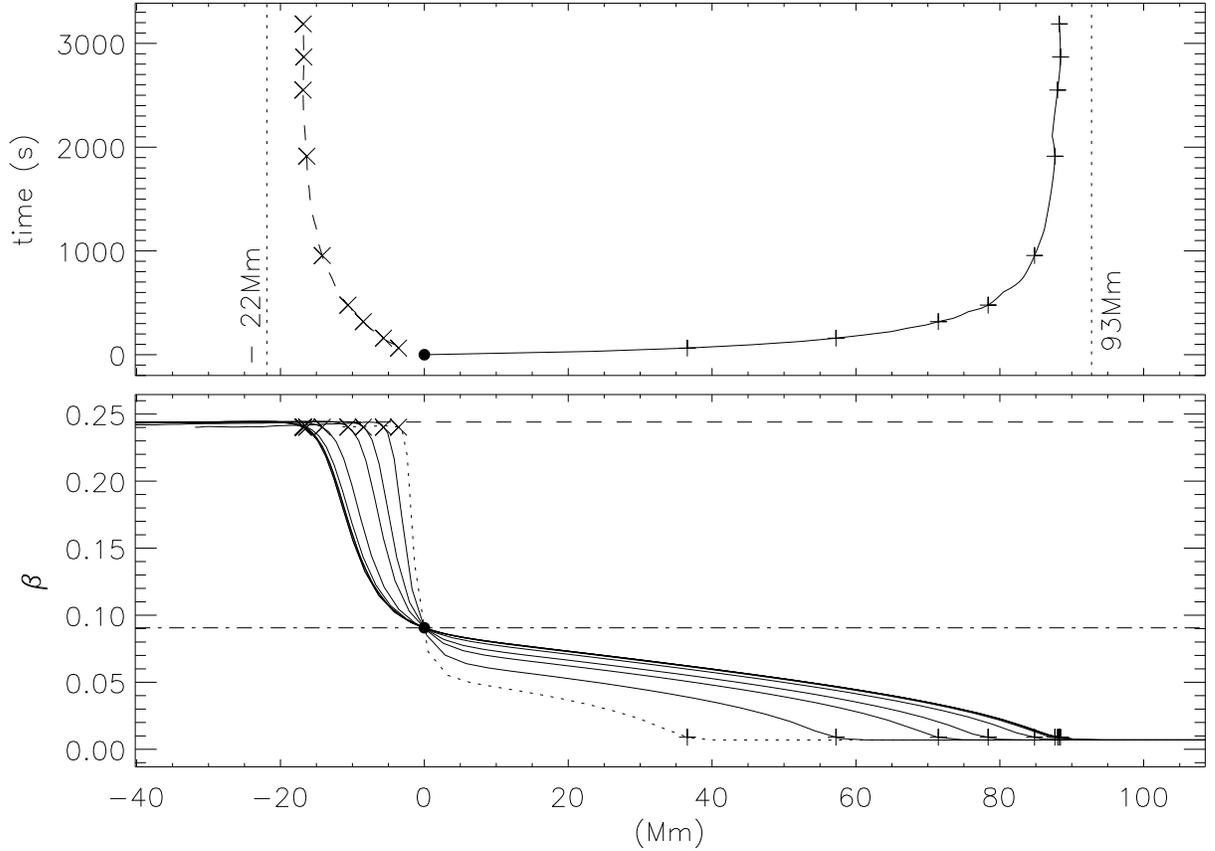}
   \caption{ Internal shock structure in the reference frame of the shock, for the simulation presented in Section \ref{sec:Simulations}. The origin is the transition between the sub-shock and the heat front, indicated with a solid circle. Both panels share the same x-axis. Bottom panel: plasma-$\beta$ as function of position in the tube for different times ($64$ s, $159$ s, $319$ s, $478$ s, $956$ s, $1913$ s, $2550$ s, $2869$ s, and $3188$ s). The dotted line represents the earliest time. Horizontal dotted line represents the value of $\beta_{0}$ that corresponds to temperature $T_{1}$, velocity $u_{0}$, and density $\rho_{0}$. Horizontal mixed dotted and dashed line represent the theoretical post-shock values. Plus signs indicate the position of the heat front (they correspond to small vertical tick marks in top panel of Fig. \ref{fig:profile_T_Beta}). Crosses indicate the position of the isothermal sub-shock (they correspond to large tick marks top panel of Fig. \ref{fig:profile_T_Beta}). Top panel: Temporal evolution of the heat front (solid line) and sub-shock (dashed line) lengths. Plus signs and crosses indicate the same positions as in the bottom panel. Leftmost vertical dotted line is the theoretical value for the length of the sub-shock calculated from equation \eqref{eqn:sub_shock_length}, and rightmost vertical dotted line is the theoretical value for the length of the heat front.} 
   \label{fig:shoulder_zoom}
\end{figure}

In the bottom panel of Fig. \ref{fig:shoulder_zoom}, the evolution of the plasma-$\beta$ of the tube is shown (in the reference frame of the shock). The thermal front extends to the right, and the isothermal sub-shock extends to the left. The top panel shows the evolution of the length of each region. 

The length (unitless) of the isothermal sub-shock, $\Delta l_{\hbox{SS} (1 \rightarrow 0)}$, is calculated replacing $T = T_{1}$ in Eq. \eqref{eqn:Int_Str_mom}, then
%
%
\begin{eqnarray}
   \label{eqn:sub_shock} 
      \Delta l_{\hbox{SS} (1 \to 0)} & = & \frac{\eta(T_{1})}{\rho_{2}u_{2}} \int_{u_{1}}^{u_{0}} \frac{u \hbox{ }du}{(u-u_{1})(u-u_{0})}, 
\end{eqnarray}
%
where integration is calculated from velocity $u_{1}$ to velocity $u_{0}={T_{1}}/{u_{1}}=2u^{*}-u_{1}$ (this latter speed corresponds to the velocity at the transition between the sub-shock and the heat front as shown in Fig. \ref{fig:fixed_point}). The plasma-$\beta$ at this point can be calculated as $\beta_{0}=\beta_{2} u_{1} {u_{2}}/{T_{2}}$. In the bottom panel of Fig. \ref{fig:shoulder_zoom}, the horizontal dotted line represents $\beta_{0}$. This value is used to approximate the position of the transition between the sub-shock and the heat front (zero horizontal coordinate in this Figure). 

The integrand in Eq. \eqref{eqn:sub_shock} diverges at both limits, therefore we will slightly modify them. We will define a smaller centered interval corresponding to a $95$\% of the original integration interval, and the integration will be performed with new limits. Defining a small interval $u_{int}= \left({u_{0}}/{u_{1}}-1 \right)/40$, the analytical solution for the length of the sub-shock is prescribed as
%
\begin{eqnarray}
   \label{eqn:sub_shock_length} 
      \Delta l_{\hbox{SS}} & = &  \frac{L^{\hbox{ion}}(T_{2})}{\sqrt{2 \gamma} M_{2,s}} \left. \left[ \frac{\log(u-1) -  \frac{u_{0}}{u_{1}} \log(|u-\frac{u_{0}}{u_{1}}|) }{\frac{u_{0}}{u_{1}}-1} \right] \right |^{\frac{u_{0}}{u_{1}} - u_{int}}_{1+u_{int}} \times 
\left\{ 
  \begin{array}{c l}
  1, & (\eta \sim T^{0}), \\
  \left( \frac{T_{1}}{T_{2}} \right)^\frac{5}{2},  & (\eta \sim T^{\frac{5}{2}}).
\end{array}
\right. 
\end{eqnarray}
%
%
where $L^{\hbox{ion}}(T_{2}) = \eta(T_{2})/\rho_{2} v_{\hbox{th}}(T_{2}) L_{0}$ is the unit-less mean-free-path of the ions at temperature $T_{2}$ (all the lengths are non-dimensionalized to characteristic length $L_{0}$); and $v_{\hbox{th}}(T_{2})$ is the thermal speed of the ions at $T_{2}$. For the simulation presented in Section \ref{sec:Simulations}, $L^{\hbox{ion}}(T_{2}) \simeq$ \protmfpunitless. 

Expression \eqref{eqn:sub_shock_length} presents values for two cases, as shown in the last part of the equation. The first one corresponds to temperature independent $\eta$, and the second one corresponds to $\eta \sim T^{\frac{5}{2}}$. The length of the sub-shock is considerably different depending on which model is used for viscosity. For the simulation we presented in the last section (temperature dependence of viscosity given by Eq. [\ref {eqn:transp_coeff_eta}]), the ratio between post-shock and pre-shock temperatures was almost one order of magnitude, therefore the sub-shock length would be approximately $10^{5/2}$ times the one predicted by a constant viscosity model. 

For the parameters of our simulation, the bottom part of Eq. \eqref{eqn:sub_shock_length} gives a length of the sub-shock approximately equal to \LSubShock, which is in good agreement with our simulation (see top panel of Fig. \ref{fig:shoulder_zoom}). In this Figure, the length of the sub-shock (negative positions) increases up to a steady-state value well approximated by the analytical solution for Prandtl number equal to zero.

\begin{figure}
  \centering
   \includegraphics{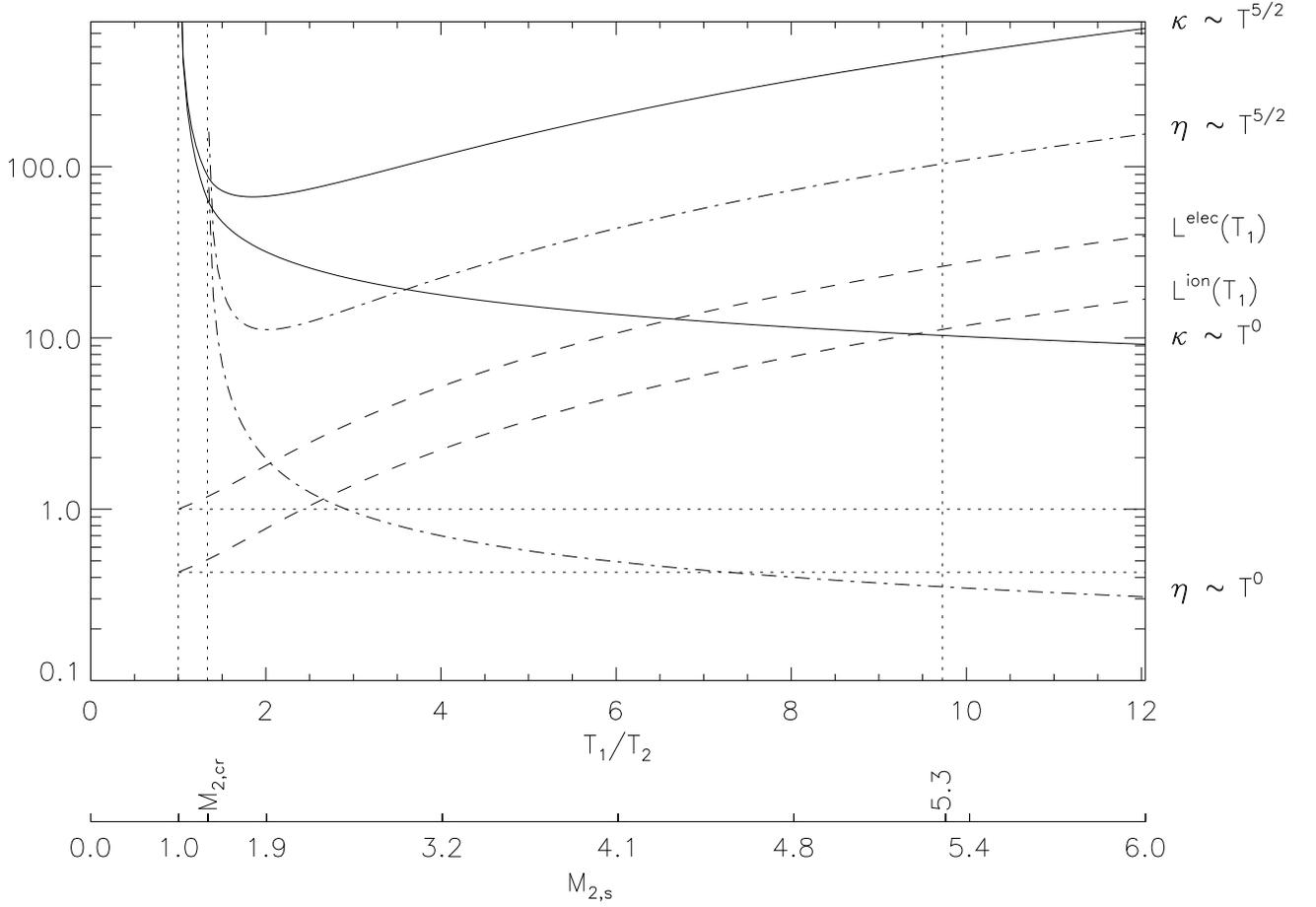}
   \caption{ Different lengths scaled with the pre-shock electron mean-free-path $L^{elec}(T_{2})$, as function of the ratio of the post-shock to the pre-shock temperature. The second axis at the bottom shows the corresponding Mach number $ M_{2,s}$. Vertical dotted lines represent cases $M_{2,s} =1$ (no shock), $M_{2,s} = M_{2,cr}$ (critical case), and $M_{2,s} =$ \MachSimShock \ (shock strength for the simulation presented in Section \ref{sec:Simulations}). The lengths in  this figure correspond to the same Reynolds and Prandtl number used for the simulation presented in Section \ref{sec:Simulations}. Bottom horizontal dotted line represents the ratio of the ion to electron mean free path at $T_{2}$, and top horizontal dotted line is the unity (electron mean free path at $T_{2}$). Solid lines depict the length of the heat front $\Delta l_{\hbox{HF}}$ (calculated from Eq. [\ref{eqn:H_F_length_const}] and [\ref{eqn:heat_front}]), for the cases labeled to the right. Mixed dotted and dashed lines represent the length of the isothermal sub-shock $\Delta l_{\hbox{SS}}$ (calculated from Eq. [\ref{eqn:sub_shock_length}]), for the same cases, and dashed lines show the mean free path of electrons and ions at post-temperature $T_{1}$.}
   \label{fig:internal_lengths}
\end{figure}

When a constant viscosity model is used, the length of the sub-shock is smaller than the mean free path of the ions for temperature ratios larger than approximately $7$ (see Fig. \ref{fig:internal_lengths}). This is not true for the case of more realistic temperature dependencies like the one used in our simulation ($\eta \sim T^{\frac{5}{2}}$). For this case, the length of the sub-shock is always more than one order of magnitude that of the ion mean-free-path, supporting the use of fluid equations. The length of the sub-shock diverges when $u_{0} = u_{1}$ (critical case), as shown in Fig. \ref{fig:internal_lengths} (mixed dotted and dashed lines), and there is no sub-shock for Mach numbers below the critical one. 

To determine the length of the heat front, we will consider the portion of $g(T,u) = 0$ between velocities $u_{0}$ and $u_{2}$, and replace the corresponding expression for the velocity $ u = u^{*}\left(1 + \sqrt{1-\frac{T}{(u^{*})^{2}}} \right) $ in Eq. \eqref{eqn:Int_Str_ene}. If we define a unitless variable $\tau = \sqrt{1-T/(u^{*})^{2}}$, and integral limits $\tau_{2}(M_{2,s}) = \sqrt{1-\frac{T_{2}}{(u^{*})^{2}}}$, and $\tau_{1}(M_{2,s}) = \sqrt{1-\frac{T_{1}}{(u^{*})^{2}}}$, the integral equation for the (unitless) thickness of the heat front becomes
%
\begin{eqnarray}
   \label{eqn:heat_front} 
      \Delta l_{\hbox{HF} (0 \to 2)} & = & \frac{4}{\rho_{2}u_{2}} \left(\frac{\gamma-1}{\gamma+1} \right) \int_{\tau_{1}}^{\tau_{2}} \frac{\kappa(\tau) \tau \hbox{ }d \tau}{(\tau \pm \tau_{1})(\tau-\tau_{2})}, 
\end{eqnarray}
%
%
where the ``$+$'' sign in the denominator corresponds to super-critical cases, and the ``$-$'' sign corresponds to the sub-critical cases. In the sub-critical cases the above integral diverges at both integral limits, and for super-critical cases, only at $\tau_{2}$. The critical case corresponds to $\tau_{1}=0$, and the integral is continuous at that point. To define a unique length for all the cases, we will integrate expression \eqref{eqn:heat_front} in a centered interval corresponding to a $95$\% of the original integration interval by defining a small interval $\tau_{int} = (\tau_{2}-\tau_{1})/40$, analogous to the one use for the isothermal sub-shock. 

The case of temperature independent thermal conductivity can be easily integrated since $\kappa(\tau)$ is a constant and can be taken outside the integral. The analytical solution for this case is 
%
%
\begin{eqnarray}
   \label{eqn:H_F_length_const} 
      \Delta l_{\hbox{HF}} & = & -4 \left(\frac{\gamma-1}{\gamma+1} \right) \sqrt{\frac{m_{p}}{m_{e}}} \frac{L^{\hbox{elec}}(T_{2})}{\sqrt{2 \gamma} M_{2,s}} \left. \left[ \frac{ \tau_{1} \log(\tau \pm \tau_{1}) \pm \tau_{2} \log(\tau_{2}-\tau)}{\tau_{1} \pm \tau_{2}} \right] \right |^{\tau_{2}-\tau_{int}}_{\tau_{1}+\tau_{int}}, 
\end{eqnarray}
%
%
Here, $L^{\hbox{elec}}(T_{2}) = \kappa(T_{2})/\rho_{2} v_{\hbox{th}}(T_{2}) L_{0}$ is the unit-less mean-free-path of the electrons at pre-shock temperature $T_{2}$, and $m_{p}$ and $m_{e}$ represent the proton and electron masses, respectively. For the simulation presented in Section \ref{sec:Simulations}, $L^{\hbox{elec}}(T_{2}) \simeq$ \elecmfpunitless. 

The temperature dependent thermal conductivity case has a slightly more complicated integral where $\kappa(\tau) = \kappa_{c} |u^{*}|^5 (1 - \tau^{2})^{\frac{5}{2}}$. For infinitely strong shocks ($M_{2,s} = \infty$), expression \eqref{eqn:heat_front} has a simpler form since $\tau_{1}(\infty) = \frac{3-\gamma}{\gamma+1}$ and $\tau_{2}(\infty) = 1$ (the divergence at $\tau_{2}$ is removed). In this case the integral converges to a finite number, and we see that the length of the heat front is $\infty$ since the factor $|u^{*}|^5/\rho_{2}u_{2}$ (that can be taken outside the integral) increases with the Mach number of the shock. The opposite occurs for the constant thermal conduction case.

For finite shocks, Eq. \eqref{eqn:heat_front} can be integrated analytically or numerically. For the realistic coronal parameters used in our simulation, the length of the heat front is approximately equal to \LHeatFront, in good agreement with our simulation (see top panel of Fig. \ref{fig:shoulder_zoom}). This length is of the order of the length of a typical coronal loop, and more than one order of magnitude the length of the mean-free-path of the electrons at the post-shock temperature. Therefore, for a real coronal situation, the heat front will probably never reach its steady-state value since it will arrive at the end-points before completing its development. At one Alfv\'{e}n time (\realalfvtime), the length of the heat front is already approximately \LHeatFrontAlfven\ long (the evolution of the heat front is shown in the top panel of Figure \ref{fig:shoulder_zoom}). 

\subsection{Justifying a Fluid Treatment}

In Fig. \ref{fig:internal_lengths}, the lengths of the isothermal sub-shock and heat front for temperature independent and temperature dependent models are plotted as function of the post-shock to pre-shock temperature ratio. All the lengths are scaled to the electron mean-free-path at temperature $T_{2}$. Solid lines depict the length of the heat front, mixed dotted and dashed lines show the length of the isothermal sub-shock, and dashed lines display the particle's mean-free-path at $T_{1}$. 

Despite its complicated analytical form, the length of the heat front for the temperature dependent case follows a simple curve that for relatively strong shocks has a shape similar to the electron mean-free-path at $T_{1}$ (dashed line). This gives some insight into the physics involved. The electron mean free path increases with post-shock temperature, as $T^{2}$ (Coulomb potential), while the thermal conduction also increases with post-shock temperature as $T^\frac{5}{2}$. The interplay between particle's mean-free-path and transport coefficient's temperature dependence determines the thickness of the shock. Whenever the diffusive coefficient's growth with temperature is faster than the particle's mean free-path, the thickness of the shock is going to be larger than the particle's mean free path. Heat is transported larger distances ahead of the shock as the temperature ratio increases, maintaining a heat front length always larger than the mean free-path. A similar analysis can be done for the sub-shock and the ion mean-free-path. When temperature independent transport coefficients are used, the results are several orders of magnitude smaller than the corresponding mean free paths, which in the past created some concern about the use of fluid equations for shock studies. 

The validity of assuming fluid equations, as opposed to kinetic theory, can also be a concern if the ratio between the diffusion heat flux and the value for streaming particles is larger than a certain value \citep{Campbell_1984}. We tested the collisionality of the plasma in our simulation, and obtained ratios smaller than $0.004$. This, and the above paragraph  conclusion, support the use of continuum fluid equations. 

For the short times involved in a flare, the shocks may not achieve their state state. The evolution of the isothermal sub-shock and the thermal front toward this state can be approximated by an exponential (see top panel of Fig. \ref{fig:shoulder_zoom}). With this approximation, the characteristic time for the development of the sub-shock is approximately equal to \SStime, and its initial (maximum) speed is less than $10$\% that of the ion thermal speed. The characteristic time for the heat front is of the order of \HHtime\ (four times the Alfv\'{e}n time), and its initial development speed is $3$\% of the electron thermal speed. We have tested the speed of the thermal front for a range of reconnection angles between $\zeta = 20$ degrees to $\zeta = 45$ degrees. For all these cases, this speed is between $3$\% to $5$\% that of the electron thermal speed. 

The electron number density and temperature evolution toward the steady state of our simulation is shown in Fig. \ref{fig:shock_heating_cooling}. The electron number density approaches its steady state from above the Rankine-Hugoniot value (top panel), and temperature approaches from below (middle panel). A short time after reconnection, the two colliding fluids accelerated at the bends get over-compressed at the center of the tube (generating densities well above the expected value  for infinitely strong shocks), and they relax towards the equilibrium value. From the top panel of the Figure, it is evident that most of the compression occurs at the sub-shock. The middle panel shows that most of the increase in temperature occurs at the thermal front. From the bottom panel of Fig. \ref{fig:shoulder_zoom}, we see that pressure achieves its steady state in a characteristic time much smaller than the ones for density and temperature. 

The bottom panel of Fig. \ref{fig:shock_heating_cooling} shows the volumetric heating rate. Most of the cooling occurs at the sub-shock where the plasma is isothermally compressed. Some cooling and heating occur at the thermal front. In this region, the plasma increases its temperature without considerably changing its density. In the sub-shock region, the velocity changes are large (see Fig. \ref{fig:fixed_point}), and it would be expected to have a strong positive contribution to the heating, however the cooling and heating are dominated by thermal conduction since the Prandtl number is small. The cooling comes from the negative curvature of the temperature to the power of $7/2$ (not shown in the Figure) as described in Eq. \eqref{eqn:heating_uniform}. Even for a small temperature gradient in the sub-shock, the heating is negative. Without viscosity, however, it is not possible to develop the needed sub-shock to connect the pre-shock state to the post-shock state. On the other hand, it is possible to only have viscosity and no thermal conduction (limit of Prandtl number equal to infinity).

%
%
\section{Conclusions}
   \label{sec:conclusions}

We have presented new general fluid equations that describe the dynamics of thin flux tubes. These tubes are assumed to be isolated from their surroundings, except by total pressure balance with their exterior (their small local radius permits almost instantaneous equilibrium with the surroundings via fast-magnetosonic waves). This assumption neglects any drag force due to the external fluid. The ideal part of these equations is applicable to a wide range of conditions for external magnetic fields and pressure. We also proposed realistic field aligned diffusive terms for viscosity and thermal conduction, valid for strongly magnetized plasmas, that include their strong dependence on temperature ($\sim T^\frac{5}{2}$). 

We solve these equations numerically for the reconnection dynamics of a thin flux tube embedded in an uniform background with small plasma-$\beta$ ($\beta =$ \realbeta). This tube was assumed to have been reconnected at an impulsive, small patchy region in a current sheet with skewed magnetic fields. Due to magnetic tension, the tube retracts at Alfv\'enic speeds reducing its length. Realistic coronal parameters (large Reynolds number and small Prandtl number) were used in the simulations. 





As the tube evolves, the initially stationary plasma is deflected at the tube bends (rotational discontinuities) and accelerated to super-sonic speeds toward the center of the tube where they collide. The small, but non-zero thermal pressure at the center of the tube stops the inflows, and the two colliding flows generate gas-dynamic shocks that move outward from the center of the tube. This shows the importance of including the parallel thermal pressure gradient in the thin flux tube equations since there is no magnetic force in that direction. The magnitude of the magnetic field remains constant across the shocks (these shocks differ from the Petschek case since they are not switch-off shocks). The presence of diffusive terms in the thin flux tube equations provides a mechanism for the shock development.

In the simulation, temperature rises almost one order of magnitude from the pre-shock value (\realTemp) to the post-shock value ($\sim$ 10MK) across the gas dynamics shocks. Nevertheless, only a small fraction (less than 10 \%) of the available magnetic energy is converted to thermal energy. The magnetic energy of the tube decreases with time as the tube shortens, and is mostly converted to kinetic energy at the bends. This suggest that the rate of the reconnection mechanism has nothing to do with the final temperature achieved. This final value only depends on the strength of the shock determined by the initial angle between the field lines. 


The inner structure of the shock is related to the transport coefficients and strength of the shock (Mach number). For solar coronal conditions (strong shocks at low Prandtl numbers), the inner structure of the shocks present two sub-regions, a heat front where most of the temperature increase occurs (its width determined by thermal conduction), and an isothermal sub-shock where the fluid is compressed (its width depends on the viscosity). For these coronal conditions, the existence of a shock transition with thermal conduction alone is not thermodynamically permissible. We estimate the thickness of these regions assuming zero Prandtl number and solving the governing differential equations analytically. The thicknesses of the sub-shock and heat front are many times the mean-free-path of the ions and electrons, respectively. This result, and our calculations regarding the free-streaming heat flux, lead us to conclude that the plasma is sufficiently collisional to use fluid equations. 


The higher density post-shock region should be readily observable. Post-shock density is almost four times greater than the initial background density, leading to sixteen times more emission measure (emission measure scales as the square of the density). The size of this region increases at a rate equal to the speed of the shock (Eq. [\ref{eqn:shock_vs}]), and at one Alfv\'{e}n time would be of the order of $10$ Mm. There are several types of observations showing bright features of this scale \citep{Masuda_1994,Tsuneta_1997_II,Warren_1999,Jiang_2006}. The radius of our tube is assumed small compared with this length, suggesting that any observable features will consist of multiple reconnection events in a small region, and several thin flux tubes being retracted outward from that region.

The jump condition for thermal pressure is achieved quite rapidly. On the other hand, density and temperature achieve their steady-state values on characteristic times longer than that for thermal pressure. The latter approaches from lower values than the corresponding Rankine-Hugoniot post-shock temperature, and the former from higher values. For large Mach numbers (large initial reconnection angles), this temporary density overshoot may be larger than the superior limit for the Rankine-Hugoniot post-shock density ratio ($\frac{\rho_{1}}{\rho_{2}} = \frac{\gamma + 1}{\gamma-1}$). The value of this ratio may exceed one order of magnitude which would manifest as even stronger X-ray emissions near the reconnection site, that would decrease in intensity as the tube evolves. Most of the cooling occurs in the isothermal sub-shock region, and most of the heating in the thermal front. 

For real coronal situations, the steady-state thickness of the shock might never be achieved since the shock will arrive at the end-points before completing its development (the steady-state thickness value is of the order of the length of the loop). For a wide range of initial pre-reconnection angles (we tested angles ranging from $\zeta = 20$ degrees to $\zeta = 45$ degrees) the speed at which the heat front develops is between $3$\% to $5$\% that of the post-shock thermal electron speed. The heat front barely compresses the plasma, since most of the compression occurs in the isothermal region. The speed at which these fronts move along the legs of the tube depends on the reconnection angle. For a magnetic field of $B_{0} =$ \realB, and electron number density $n_{0} =$ \realnumdens, the range in speed for the thermal fronts is between 
$450$ km s$^{-1}$ to $650$ km s$^{-1}$ for angles between $\zeta = 20$ degrees to $\zeta = 45$ degrees. For larger angles, it is possible for the heat front to overcome the bends.

\begin{figure}
   \centering
   \includegraphics{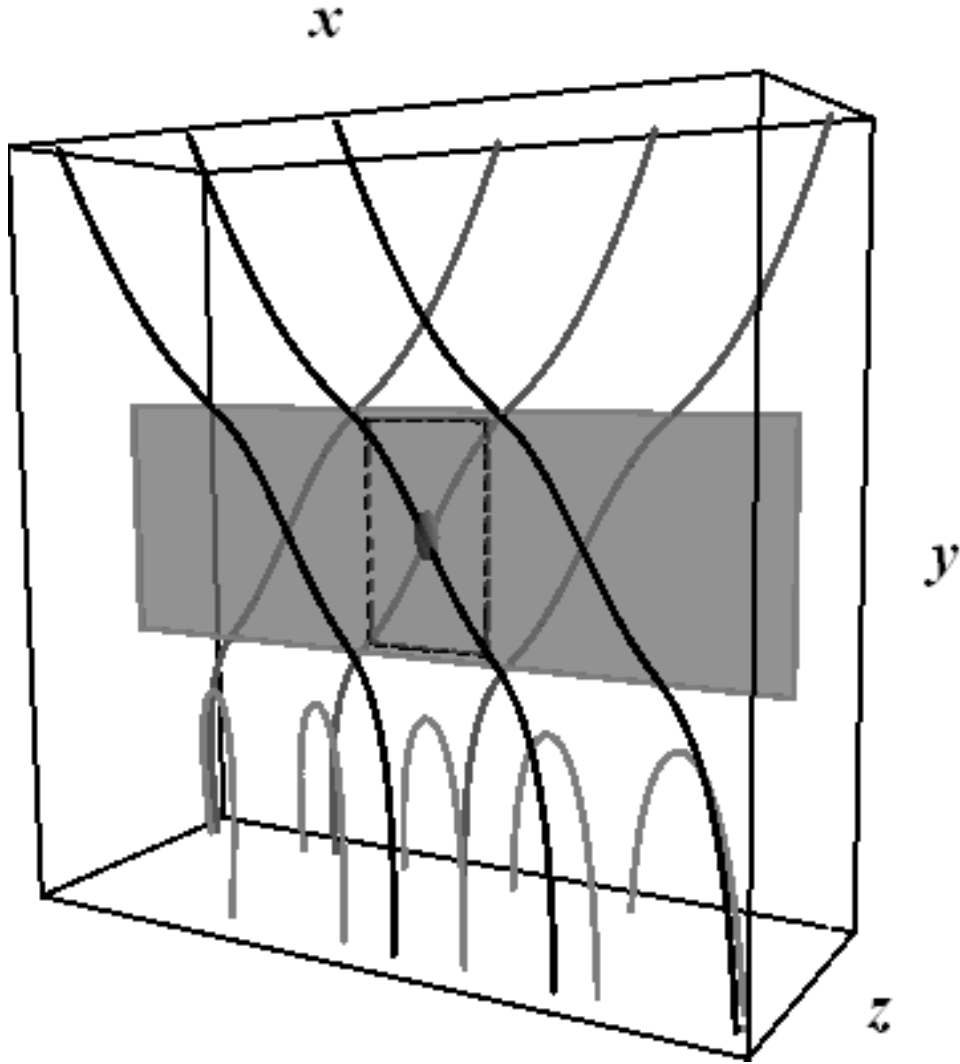}
   \caption
   {
      Flare current sheet cartoon. The solid rectangle represents the current sheet plane located above the post-flare arcade (gray parabolic loops at the bottom). The black lines depict some magnetic field lines on the side of the current sheet that is closer to the viewer, and the dark gray ones correspond to field lines on the back side of the current sheet. The lower plane of the box represents the solar surface. The small sphere in the current sheet shows a patchy reconnection region. Only a small bundle of field lines that intersect this region are going to reconnect. The dashed rectangle in the plane of the current sheet corresponds to the region shown in Fig. \ref{fig:current_sheet}. }
   \label{fig:flare_CS} 
\end{figure}

Heat fronts are possibly relevant for chromospheric evaporation in flares. This is not directly treatable in our model since it is confined to the current sheet whose edge lays high above the chromosphere (see Fig \ref{fig:flare_CS}). We expect the heat fronts found in our model to continue along the field lines towards their footpoints in the chromosphere. The temperature at the heat front increases approximately up to the post-shock Rankine-Hugoniot value, which could be one order of magnitude larger than the initial temperature of the loop. At these high temperatures chromospheric material can be evaporated, and X-ray emissions from the foot points might be observed. The upward flow of evaporated chromospheric material would eventually encounter the sub-shock that is descending along the legs of the tube. What would happen after is a subject for further investigation.

We neglected radiation in our model. The radiative cooling time at the pre-reconnection initial conditions is several orders of magnitude larger than the Alfv\'{e}n time. The density increase by as much as a factor of four would by itself decrease the radiative cooling time. Temperature, however, also increases one order of magnitude which dramatically increases the cooling time since the radiative loss function decreases with increasing temperature at temperatures higher than $1$ MK, \citep{Cook_1989,Martens_2000,Rosner_1978}. Therefore, radiation is a negligible effect in the reconnection process.

Our model assumed a very simple initial geometry: uniform background conditions with pre-reconnection field lines that form an angle at both sides of the reconnection. In future work, we will investigate a more realistic model with, for example, a Green-Syrovatski{\v i} \citep{Green_1965,Syrovatskii_1971} current sheet with a guide field. This model presents new phenomena as the magnetic field magnitude changes along the tube. 


\acknowledgments

This work was supported by NSF and NASA.  We thank Terry Forbes for useful discussions.

The research in this paper is part of a dissertation by S. Guidoni to be submitted to the Graduate School at the Montana  State University in partial fulfillment of the requirements for completion of a Ph.D. degree.

%




\begin{thebibliography}{66}
\expandafter\ifx\csname natexlab\endcsname\relax\def\natexlab#1{#1}\fi

\bibitem[{{Biernat} {et~al.}(1987){Biernat}, {Heyn}, \&
  {Semenov}}]{Biernat_1987}
{Biernat}, H.~K., {Heyn}, M.~F., \& {Semenov}, V.~S. 1987, \jgr, 92, 3392

\bibitem[{{Biernat} {et~al.}(1998){Biernat}, {Semenov}, \&
  {Rijnbeek}}]{Biernat_1998}
{Biernat}, H.~K., {Semenov}, V.~S., \& {Rijnbeek}, R.~P. 1998, \jgr, 103, 4693

\bibitem[{{Birn} {et~al.}(2001){Birn}, {Drake}, {Shay}, {Rogers}, {Denton},
  {Hesse}, {Kuznetsova}, {Ma}, {Bhattacharjee}, {Otto}, \&
  {Pritchett}}]{Birn_2001}
{Birn}, J., {et~al.} 2001, \jgr, 106, 3715

\bibitem[{{Biskamp}(1986)}]{Biskamp_1986}
{Biskamp}, D. 1986, Physics of Fluids, 29, 1520

\bibitem[{Biskamp \& Schwarz(2001)}]{Biskamp_2001}
Biskamp, D., \& Schwarz, E. 2001, Physics of Plasmas, 8, 4729

\bibitem[{{Braginskii}(1965)}]{Braginskii_1965}
{Braginskii}, S.~I. 1965, Reviews of Plasma Physics, 1, 205

\bibitem[{{Campbell}(1984)}]{Campbell_1984}
{Campbell}, P.~M. 1984, \pra, 30, 365

\bibitem[{{Cargill} {et~al.}(1995){Cargill}, {Mariska}, \&
  {Antiochos}}]{Cargill_1995}
{Cargill}, P.~J., {Mariska}, J.~T., \& {Antiochos}, S.~K. 1995, \apj, 439, 1034

\bibitem[{{Cook} {et~al.}(1989){Cook}, {Cheng}, {Jacobs}, \&
  {Antiochos}}]{Cook_1989}
{Cook}, J.~W., {Cheng}, C., {Jacobs}, V.~L., \& {Antiochos}, S.~K. 1989, \apj,
  338, 1176

\bibitem[{{Coroniti}(1970)}]{Coroniti_1970}
{Coroniti}, F.~V. 1970, Journal of Plasma Physics, 4, 265

\bibitem[{{Courant} {et~al.}(1967){Courant}, {Friedrichs}, \&
  {Lewy}}]{CFL_1967}
{Courant}, R., {Friedrichs}, K., \& {Lewy}, H. 1967, IBM Journal, 215

\bibitem[{{Craig}(2008)}]{Craig_2008}
{Craig}, I.~J.~D. 2008, \aap, 487, 1155

\bibitem[{{Craig} \& {Litvinenko}(2009)}]{Craig_2009}
{Craig}, I.~J.~D., \& {Litvinenko}, Y.~E. 2009, \aap, 501, 755

\bibitem[{{Craig} {et~al.}(2005){Craig}, {Litvinenko}, \&
  {Senanayake}}]{Craig_2005}
{Craig}, I.~J.~D., {Litvinenko}, Y.~E., \& {Senanayake}, T. 2005, \aap, 433,
  1139

\bibitem[{{Forbes}(2000)}]{Forbes_2000}
{Forbes}, T.~G. 2000, \jgr, 105, 23153

\bibitem[{{Forbes} \& {Malherbe}(1986)}]{Forbes_1986}
{Forbes}, T.~G., \& {Malherbe}, J.~M. 1986, \apjl, 302, L67

\bibitem[{{Forbes} {et~al.}(1989){Forbes}, {Malherbe}, \&
  {Priest}}]{Forbes_1989}
{Forbes}, T.~G., {Malherbe}, J.~M., \& {Priest}, E.~R. 1989, \solphys, 120, 285

\bibitem[{{Fowles}(1975)}]{Fowles_1975}
{Fowles}, G.~R. 1975, Physics of Fluids, 18, 776

\bibitem[{{Gilbarg} \& {Paolucci}(1953)}]{Gilbarg_1953}
{Gilbarg}, D., \& {Paolucci}, D. 1953, Indiana Univ. Math. J., 2, 617

\bibitem[{{Goldstein}(1959)}]{Goldstein_Cl_Mech}
{Goldstein}, H. 1959, {Classical mechanics} (Addison-Wesley Publishing Company,
  Inc., Reading, Mass., U.S.A, London, England)

\bibitem[{{Grad}(1952)}]{Grad_1952}
{Grad}, H. 1952, Communications on pure and applied mathematics, v, 257

\bibitem[{{Green}(1965)}]{Green_1965}
{Green}, R.~M. 1965, in IAU Symposium, Vol.~22, Stellar and Solar Magnetic
  Fields, ed. {R.~Lust}, 398--+

\bibitem[{{Heyn} {et~al.}(1988){Heyn}, {Biernat}, {Rijnbeek}, \&
  {Semenov}}]{Heyn_1988}
{Heyn}, M.~F., {Biernat}, H.~K., {Rijnbeek}, R.~P., \& {Semenov}, V.~S. 1988,
  Journal of Plasma Physics, 40, 235

\bibitem[{{Hugoniot}(1887)}]{Hugoniot_1887}
{Hugoniot}, H. 1887, Journal de l'Ecole Polytechnique, 57, 3

\bibitem[{{Jiang} {et~al.}(2006){Jiang}, {Liu}, {Liu}, \&
  {Petrosian}}]{Jiang_2006}
{Jiang}, Y.~W., {Liu}, S., {Liu}, W., \& {Petrosian}, V. 2006, \apj, 638, 1140

\bibitem[{{Kennel}(1988)}]{Kennel_1988}
{Kennel}, C.~F. 1988, \jgr, 93, 8545

\bibitem[{{Klimchuk}(2001)}]{Klimchuk_2001}
{Klimchuk}, J.~A. 2001, Space Weather (Geophysical Monograph 125), ed.~P.~Song,
  H.~Singer, G.~Siscoe (Washington: Am.~Geophys.~Un.), 143 (2001), 125, 143

\bibitem[{{Kopp} \& {Pneuman}(1976)}]{Kopp_1976}
{Kopp}, R.~A., \& {Pneuman}, G.~W. 1976, \solphys, 50, 85

\bibitem[{{Landau} \& {Lifshitz}(1959)}]{Landau_1959}
{Landau}, L.~D., \& {Lifshitz}, E.~M. 1959, {Fluid mechanics} (Oxford: Pergamon
  Press, 1959)

\bibitem[{{Lin} {et~al.}(2008){Lin}, {Cranmer}, \& {Farrugia}}]{Lin_2008}
{Lin}, J., {Cranmer}, S.~R., \& {Farrugia}, C.~J. 2008, Journal of Geophysical
  Research (Space Physics), 113, 11107

\bibitem[{{Lin} {et~al.}(2003){Lin}, {Soon}, \& {Baliunas}}]{Lin_2003}
{Lin}, J., {Soon}, W., \& {Baliunas}, S.~L. 2003, New Astronomy Review, 47, 53

\bibitem[{{Linton} \& {Longcope}(2006)}]{Linton_2006}
{Linton}, M.~G., \& {Longcope}, D.~W. 2006, \apj, 642, 1177

\bibitem[{{Litvinenko}(2005)}]{Litvinenko_2005}
{Litvinenko}, Y.~E. 2005, \solphys, 229, 203

\bibitem[{{Longcope} {et~al.}(2009){Longcope}, {Guidoni}, \&
  {Linton}}]{Longcope_2009}
{Longcope}, D.~W., {Guidoni}, S.~E., \& {Linton}, M.~G. 2009, \apjl, 690, L18

\bibitem[{{Low}(2001)}]{Low_2001}
{Low}, B.~C. 2001, \jgr, 106, 25141

\bibitem[{{Martens} {et~al.}(2000){Martens}, {Kankelborg}, \&
  {Berger}}]{Martens_2000}
{Martens}, P.~C.~H., {Kankelborg}, C.~C., \& {Berger}, T.~E. 2000, \apj, 537,
  471

\bibitem[{{Masuda} {et~al.}(1994){Masuda}, {Kosugi}, {Hara}, {Tsuneta}, \&
  {Ogawara}}]{Masuda_1994}
{Masuda}, S., {Kosugi}, T., {Hara}, H., {Tsuneta}, S., \& {Ogawara}, Y. 1994,
  \nat, 371, 495

\bibitem[{{McKenzie} \& {Hudson}(1999)}]{McKenzie_1999}
{McKenzie}, D.~E., \& {Hudson}, H.~S. 1999, \apjl, 519, L93

\bibitem[{{McKenzie} \& {Savage}(2009)}]{McKenzie_2009}
{McKenzie}, D.~E., \& {Savage}, S.~L. 2009, \apj, 697, 1569

\bibitem[{{Nitta} {et~al.}(2002){Nitta}, {Tanuma}, \& {Maezawa}}]{Nitta_2002}
{Nitta}, S., {Tanuma}, S., \& {Maezawa}, K. 2002, \apj, 580, 538

\bibitem[{{Parker}(1957)}]{Parker_1957}
{Parker}, E.~N. 1957, \jgr, 62, 509

\bibitem[{{Petschek}(1964)}]{Petschek_1964}
{Petschek}, H.~E. 1964, NASA Special Publication, 50, 425

\bibitem[{{Petschek} \& {Thorne}(1967)}]{Petschek_1967}
{Petschek}, H.~E., \& {Thorne}, R.~M. 1967, \apj, 147, 1157

\bibitem[{{Priest}(1981)}]{Priest_1981}
{Priest}, E.~R. 1981, {Solar flare magnetohydrodynamics} (New York, Gordon and
  Breach Science Publishers)

\bibitem[{{Pudovkin} \& {Semenov}(1985)}]{Pudovkin_1985}
{Pudovkin}, M.~I., \& {Semenov}, V.~S. 1985, Space Science Reviews, 41, 1

\bibitem[{{Rankine}(1870)}]{Rankine_1870}
{Rankine}, W.~J.~M. 1870, Phil. Trans. R. Soc. Lond., 160, 277

\bibitem[{{Rosner} {et~al.}(1978){Rosner}, {Tucker}, \& {Vaiana}}]{Rosner_1978}
{Rosner}, R., {Tucker}, W.~H., \& {Vaiana}, G.~S. 1978, \apj, 220, 643

\bibitem[{{Russell} \& {Elphic}(1978)}]{Russell_1978}
{Russell}, C.~T., \& {Elphic}, R.~C. 1978, Space Science Reviews, 22, 681

\bibitem[{{Schettino} {et~al.}(2010){Schettino}, {Poletto}, \&
  {Romoli}}]{Schettino_2010}
{Schettino}, G., {Poletto}, G., \& {Romoli}, M. 2010, \apj, 708, 1135

\bibitem[{{Seaton} \& {Forbes}(2009)}]{Seaton_2009}
{Seaton}, D.~B., \& {Forbes}, T.~G. 2009, \apj, 701, 348

\bibitem[{{Semenov} {et~al.}(1983){Semenov}, {Heyn}, \&
  {Kubyshkin}}]{Semenov_1983}
{Semenov}, V.~S., {Heyn}, M.~F., \& {Kubyshkin}, I.~V. 1983, Soviet Astronomy,
  27, 660

\bibitem[{{Semenov} {et~al.}(1992){Semenov}, {Kubyshkin}, {Levedeva},
  {Sidneva}, {Biernat}, {Heyn}, {Besser}, \& {Rijnbeek}}]{Semenov_1992}
{Semenov}, V.~S., {Kubyshkin}, I., {Levedeva}, V.~V., {Sidneva}, M.~V.,
  {Biernat}, H.~K., {Heyn}, M.~F., {Besser}, B.~P., \& {Rijnbeek}, R.~P. 1992,
  \jgr, 97, 4251

\bibitem[{{Semenov} {et~al.}(1998){Semenov}, {Volkonskaya}, \&
  {Biernat}}]{Semenov_1998}
{Semenov}, V.~S., {Volkonskaya}, N.~N., \& {Biernat}, H.~K. 1998, Physics of
  Plasmas, 5, 3242

\bibitem[{{Sheeley} {et~al.}(2004){Sheeley}, {Warren}, \&
  {Wang}}]{Sheeley_2004}
{Sheeley}, Jr., N.~R., {Warren}, H.~P., \& {Wang}, Y. 2004, \apj, 616, 1224

\bibitem[{{Shu}(1991)}]{Shu_VolII}
{Shu}, F. 1991, {Physics of Astrophysics, Vol. II: Gas Dynamics} (University
  Science Books, 648 Broadway, Suite 902, New York, NY 10012)

\bibitem[{{Skender} {et~al.}(2003){Skender}, {Vr{\v s}nak}, \&
  {Martinis}}]{Skender_2003}
{Skender}, M., {Vr{\v s}nak}, B., \& {Martinis}, M. 2003, \pre, 68, 046405

\bibitem[{Soward(1982)}]{Soward_1982_II}
Soward, A.~M. 1982, Journal of Plasma Physics, 28, 415

\bibitem[{Soward \& Priest(1982)}]{Soward_1982}
Soward, A.~M., \& Priest, E.~R. 1982, Journal of Plasma Physics, 28, 335

\bibitem[{{Spitzer}(1962)}]{Spitzer_1962}
{Spitzer}, L. 1962, {Physics of Fully Ionized Gases} (New York: Interscience
  (2nd edition))

\bibitem[{{Spruit}(1981)}]{Spruit_1981}
{Spruit}, H.~C. 1981, \aap, 98, 155

\bibitem[{{Sweet}(1958)}]{Sweet_1958}
{Sweet}, P.~A. 1958, Il Nuovo Cimento Suppl., 8, 188–196

\bibitem[{{Syrovatski{\v i}}(1971)}]{Syrovatskii_1971}
{Syrovatski{\v i}}, S.~I. 1971, Soviet Journal of Experimental and Theoretical
  Physics, 33, 933

\bibitem[{{Thomas}(1944)}]{Thomas_1944}
{Thomas}, L.~H. 1944, \jcp, 12, 449

\bibitem[{{Tsuneta}(1997)}]{Tsuneta_1997_I}
{Tsuneta}, S. 1997, \apj, 483, 507

\bibitem[{{Tsuneta} {et~al.}(1997){Tsuneta}, {Masuda}, {Kosugi}, \&
  {Sato}}]{Tsuneta_1997_II}
{Tsuneta}, S., {Masuda}, S., {Kosugi}, T., \& {Sato}, J. 1997, \apj, 478, 787

\bibitem[{{Warren} {et~al.}(1999){Warren}, {Bookbinder}, {Forbes}, {Golub},
  {Hudson}, {Reeves}, \& {Warshall}}]{Warren_1999}
{Warren}, H.~P., {Bookbinder}, J.~A., {Forbes}, T.~G., {Golub}, L., {Hudson},
  H.~S., {Reeves}, K., \& {Warshall}, A. 1999, \apjl, 527, L121

\end{thebibliography}

\clearpage

%
%

\clearpage

%
%

\clearpage

%
%
%

\clearpage

%
%
%

\clearpage

%
%

\clearpage

%
%
%
%

\clearpage

%
%
%
%

\clearpage

%
%
%
 
\clearpage

%
%
%

\clearpage

%
%
%

\end{document}